\documentclass[prb,aps,floats,showpacs,amsmath, twocolumn, amssymb]{revtex4}

\usepackage{amssymb}
\usepackage{amsmath}
\usepackage{epsfig}
\usepackage{hyperref}
\usepackage{color}
\usepackage{xcolor} 

\newcommand{\be}{\begin{equation}}
\newcommand{\ee}{\end{equation}}
\newcommand{\bea}{\begin{eqnarray}}
\newcommand{\eea}{\end{eqnarray}}

\newcommand{\tlt}{\tilde{t}}
\newcommand{\tr}{\tilde{r}}

\newcommand{\Omegam}{\Omega_{\rm M}}

\def    \bse{\begin{subequations}}
\def    \ese{\end{subequations}}

\begin{document}

\title{Photon propagation in a one-dimensional optomechanical lattice}

\author{Wei Chen}
\affiliation{Department of Physics, McGill University, Montreal, Canada H3A 2T8 }
\author{Aashish A.\ Clerk} 
\email{clerk@physics.mcgill.ca}
\affiliation{Department of Physics, McGill University, Montreal, Canada H3A 2T8 }

\begin{abstract}
We consider a one-dimensional optomechanical lattice where each site is strongly driven by 
a control laser to enhance the basic optomechanical interaction.  
We then study the propagation of photons injected by an additional probe laser beam; this is the
lattice-generalization of the well-known optomechanically-induced transparency (OMIT)
effect in a single optomechanical cavity.  We find an interesting interplay between OMIT-type physics
and geometric, Fabry-Perot type resonances.  In particular, phonon-like polaritons can give rise to
high-amplitude transmission resonances which are much narrower than the scale set by internal photon losses.
We also find that the local photon density of states in the lattice exhibits OMIT-style interference features.  It is thus 
far richer than would be expected by just looking at the band structure of the dissipation-free coherent system.  
\end{abstract}

\maketitle

\section{Introduction}

The rapidly growing field of quantum optomechanics involves studying the coupling of mechanical motion
to light, with the prototypical structure being an optomechanical cavity:  photons in a single mode of an electromagnetic cavity interact via radiation pressure with the position
of a mechanical resonator(see Ref.~\onlinecite{Florianreview} for a recent review).  Among the many recent  breakthroughs
in this area are the laser cooling of mechanical motion to the ground state \cite{Teufel2011a,Painter2011}, the generation of squeezed light via the optomechanical interaction \cite{Brooks2012,Amir2013,Purdy2013},
and demonstration of entanglement between microwaves and mechanical motion \cite{Lehnert2013}.  

A particularly exciting direction for this field is the development of
multi-mode structures composed of many individual optomechanical cavities which are coupled via photonic and/or phononic tunnelling.  Such multi-site optomechanical arrays can be realized in so-called optomechanical crystals, fabricated using a planar photonic bandgap material, and using defects to localize both mechanical and optical modes \cite{Eichenfield2009a,Eichenfield2009b}. Recent theoretical studies have investigated synchronization and phase-transition physics in these systems  \cite{Heinrich2011,Holmes2012, Tomadin2012,Ludwig2013}, 
applications to quantum information processing \cite{Schmidt2012}, as well as the effective band structure describing excitations when the system is driven with a red-detuned laser \cite{Schmidt2013}. 
Arrays based on multiple membranes in a single cavity have also been studied \cite{Xuereb2012, Xuereb2013, Xuereb2013-2}.

\begin{figure}[ht]
\begin{center}
\vspace{0.7cm}
\includegraphics[scale=0.35]{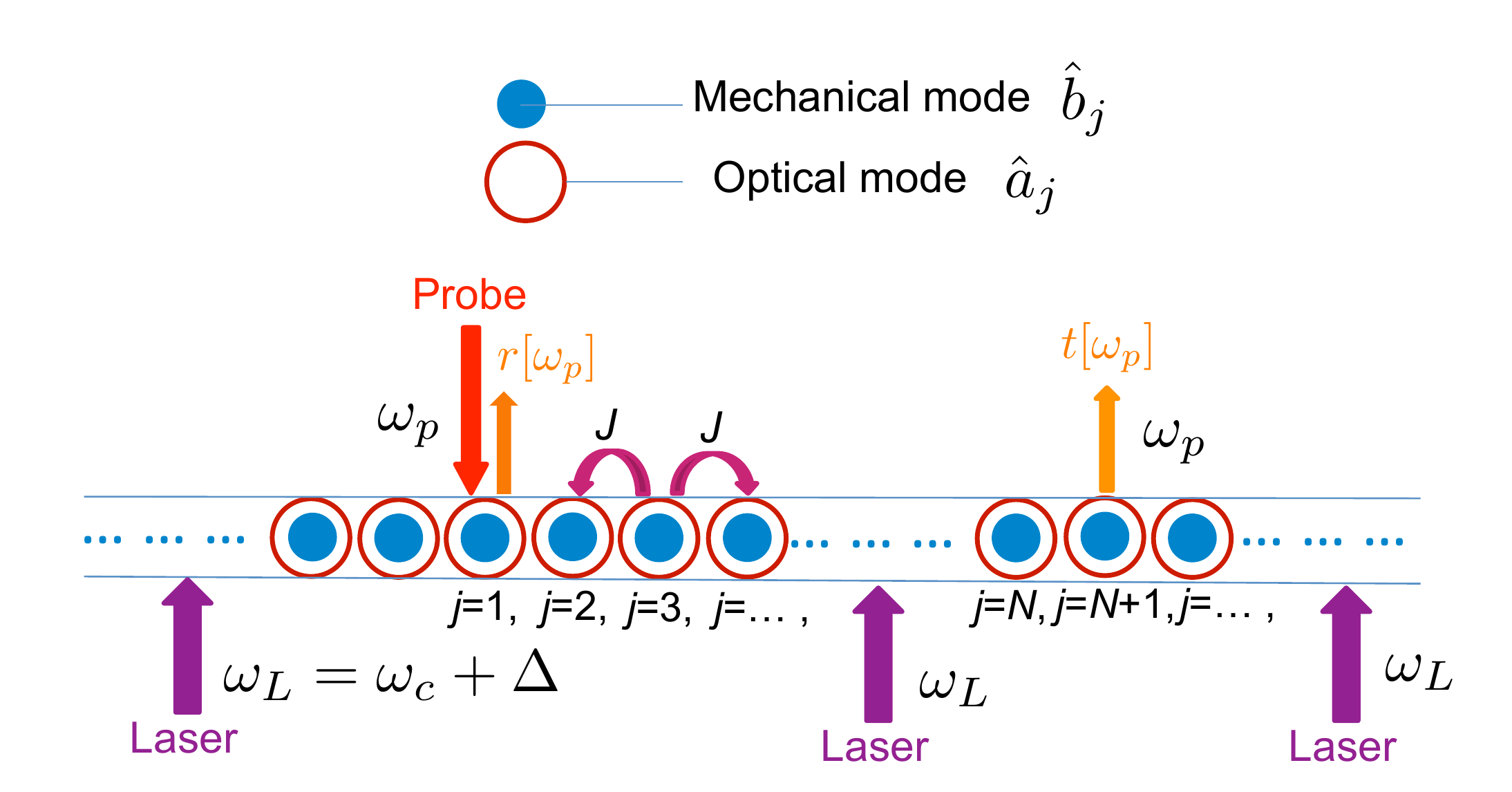}
\caption{(color online) 
Schematic picture of the 1D optomechanical array studied in  
this work. On each site, the blue (red) circle represents a localized 
mechanical  (photonic) mode; they interact locally via a standard
radiation pressure coupling. Photons can hop between neighbouring site with hopping strength $J$. A uniform control laser is applied to each site at frequency $\omega_L=\omega_c+\Delta$, where $\Delta$ is the detuning between the laser frequency and cavity photon frequency $\omega_c$.
For transport, external waveguides are attached to sites $1$ and $N+1$; these are used to send and extract
probe light at frequency $\omega_p$.  
}
\label{fig:setup}
\end{center}
\end{figure}

In this work, we consider how a prototypical optomechanical effect -- optomechanically induced transparency (OMIT) \cite{Agarwal2010,Weis2010,SafaviNature,Teufel2011}-- is modified in a lattice setting. OMIT in a single cavity is an effect whereby even a weak optomechanical interaction can, in the presence of a large-amplitude control laser, lead to a strong modification of the cavity photon density of states.  This modification can in turn be probed using either the reflection or transmission of a weak auxiliary probe field.  It is analogous to the well-known electromagnetically-induced transparency effect (EIT) \cite{Harris1990, Boller1991,Fleischhauer2005} involving the interaction of a three-level atom with light, where destructive interference prevents optical absorption by the atom.  We consider a 1D optomechanical array where photons can tunnel between adjacent sites, and consider what happens when, similar to a standard OMIT experiment, the entire system is driven by a large-amplitude control laser (detuned in such a manner to avoid any instability).  We work in the standard regime where the single-photon
optomechanical coupling is negligible, and only the drive-enhanced many-photon coupling plays a role.   Unlike the recent work from Schmidt et al. \cite{Schmidt2013}, we restrict attention to 1D, but do more than simply characterize the band structure describing excitations in this system:  we also provide a full description of transport phenomena, where additional probe photons are injected at one site of the lattice, and extracted at some other point in the lattice (see Fig.~1).  Our work also strongly differs from the previous study by Chang et al.\cite{Chang2011}, who studied an alternate geometry where a regular 1D array of optomechanical cavities is side-coupled to a one-dimensional waveguide.  In that system there is no direct tunnelling between optomechanical cavities, whereas in our system, probe photons can only be transported via such a tunnel coupling.  As we discuss, this gives rise to markedly different behaviours.

Our key results can be summarized as follows.  After introducing our system in Sec.~\ref{sec:Model}, we first look at the local photon density of states of the lattice in 
Sec.~\ref{sec:DOS}.  In the presence of a many-photon optomechanical coupling but ignoring dissipation, one will generically obtain an energy gap between the polariton bands of the coupled system.  Similar to OMIT, we find that this gap can persist even if it is much smaller than the intrinsic photon damping rate $\kappa_0$.  Unlike OMIT, the size of the spectral gap in the coupled system can be identical in size to that expected in the coherent system.  We show how this density-of-states phenomenon could be directly measured in an OMIT-style experiment involving the local reflection of probe light from a given site of the lattice.  

We next consider in Sec.~\ref{sec:Transport} the question of transport in our system, where probe light is injected at site $j=1$ and extracted at site $j= N + 1$ of the lattice.  We find that polariton modes that are almost entirely phononic can contribute strongly to photon transport.  We also study the interplay between these OMIT-style effects and the presence of geometric resonances (caused either by reflections resulting from the coupling to the external probe waveguides, or from the finite-size of the lattice).  We again find a striking result:  interference effects involving phonon-like polaritons can yield transmission resonances that have extremely high maximum transmissions, and which have a frequency width much narrower than the intrinsic photon loss rate of the lattice $\kappa_0$.  Note that the interplay between conventional atomic EIT and geometric optical resonances was previously studied by Lukin et al.\cite{Lukin1998}; as we discuss further in Sec.~\ref{subsec:EITanalogy}, there are marked differences from our work, as the analogy between OMIT in our system and propagation in an atomic EIT medium is not complete. 

We stress that the effects we describe in this work, like OMIT itself, are not simply the result of the coherent hybridization of light and matter--  the effects we describe persist even in a regime where this coherent coupling is much smaller than the photonic dissipation.  Instead, these effects rely crucially on the dissipative physics of the system, in particular the markedly different intrinsic damping rates of photons and phonons, and the fact that the dissipation is not ``diagonal" in the basis of energy eigenstates of the coherent system (the polariton basis).  
In the case of EIT, this crucial difference has been emphasized in a recent work 
by Anisimov et al \cite{Sanders2011}.



\section{Model and polariton bandstructure}
\label{sec:Model}

\subsection{Optomechanical lattice Hamiltonian}
We first consider an infinite one dimensional lattice where each site $j$ contains a localized photonic
and phononic mode (lowering operators $\hat{a}_j$, $\hat{b}_j$ respectively), coupled locally via the standard dispersive optomechanical interaction (strength $g$).  
We allow for photon hopping between neighbouring sites via standard tight-binding hopping terms (hopping strength $J$); for simplicity, we ignore 
mechanical hopping, though this could easily be included in the model. 
The Hamiltonian $\hat{H}$ of the lattice thus takes the form $\hat{H} = \hat{H}_{\rm coh} +
\hat{H}_{\rm diss}$ where
\bse
\begin{eqnarray}\label{eq:coherent} 
	&&\hat{H}_{\rm coh}=\sum_{j}\hat{H}_{j}  -J\sum_{\langle i,j\rangle}\hat{a}^\dag_i \hat{a}_j, \\ 
	&&\hat{H}_{j}=\omega_\text{c} \hat{a}^\dag_j \hat{a}_j + \Omegam 
	\hat{b}^\dag_j \hat{b}_j +g(\hat{b}_j+\hat{b}_j^{\dag})\hat{a}_j^{\dag}\hat{a}_j,
\end{eqnarray}
\ese
Here $\hat{H}_{j}$ is the standard Hamiltonian of a single optomechanical 
cell~\cite{Florianreview}, with $\omega_\text{c}$ the photon mode energy and $\Omegam$ the phonon energy.OMIT
$\hat{H}_\text{dis}$ describes both dissipation of photon and phonons.  We assume that phonons and photons on each site are coupled to their own (independent) Markovian, gaussian
dissipative baths, and treat these as is done in standard input-output theory ~\cite{Clerkreview, Zollerbook}.  This gives rise to a damping rate $\gamma$ for each phonon mode, and a damping rate $\kappa_0$ for each photon mode.  $\kappa_0$ includes both dissipation due to internal losses, as well as the coupling to port used to drive each
site of the lattice.  We will focus throughout on the resolved sideband regime here, where $\kappa_0 \ll \Omegam$.  Note that while we focus on
an infinite system, the key effects we describe also have direct relevance to finite-sized systems (see Appendix \ref{appendix:Finite}).

As mentioned in the introduction, the model described above could most easily be realized experimentally using so-called optomechanical crystals, where an array of optomechanical cells is formed by creating a superlattice of defects in a photonic bandgap material (typically a patterned silicon membrane); each defect localizes an optical and a phononic mode, providing an effective optomechanical cavity.  Fabrication and characterization of one and two dimensional optomechanical crystals have been achieved in several recent experiments\cite{PainterAPL2012, Painter2014}, where typical parameters have the values $\omega_c/2\pi \simeq 200$ THz, $\Omega_M/2\pi= 10$ GHz, $g/2\pi=250 $kHz and the intrinsic optical and mechanical damping rates are in the range $200 {\rm KHz} - 2 {\rm GHz}$ and $10 - 200 {\rm KHz}$ respectively. The photon hopping $J$ depends on the distance between neighboring sites and typically has the same order of magnitude as the photon frequency $\omega_c$ in experiments.  As discussed recently in Ref.~\onlinecite{Schmidt2013}, for such realistic optomechanical crystals, one expects the phonon hopping to be smaller than the photon hopping $J$ by a large factor $\sim 10^4$;
hence the neglect of phonon hopping in our model is well-motivated.

Note that for simplicity, in this work we do not consider the effect of disorder arising, e.g.~, from site-to-site fluctuations of the photon and/or phonon frequencies.  Such disorder effects would generically give rise to localization effects in a 1D system.  However, as discussed in Ref.~\onlinecite{Schmidt2013}, 
for realistic finite-sized cavity arrays the disorder can be weak enough such that the localization length is much larger than the size of the array (see, e.g.~Ref.~\onlinecite{Notomi2008}).

\subsection{Many-photon optomechanical coupling}

In standard single-cavity optomechanics experiments, the single photon coupling is extremely weak (i.e.~$g \ll \kappa_0$), and hence the cavity
is strongly driven to enhance the interaction.  We assume a similar situation in our lattice, now taking each site to be strongly driven by a control laser
at frequency $\omega_L$; we take the laser amplitude to be the same on all sites.  Working in a rotating frame at the laser frequency, the laser will induce 
a classical, time-independent mean amplitude for each photon mode $\bar{a}_j$ and each mechanical mode $\bar{b}_j$.  These 
stationary amplitudes are readily found from the classical equations of motion for our system:
\bse
\begin{eqnarray}
	0&=&\left(i\Delta-\frac{\kappa_0}{2}\right)\bar{a}_j -i g \bar{a}_j (\bar{b}_j+\bar{b}^*_j) \\
	&&+iJ(\bar{a}_{j-1}+\bar{a}_{j+1}) - \sqrt{\kappa_0}\bar{a}_\text{in},\nonumber\\
	0&=&\left(-i\Omegam - \frac{\gamma}{2}\right)\bar{b}_j-i g|\bar{a}^2_j|. 
\end{eqnarray}
\ese
Here, $\bar{a}_{\rm in}$ is the amplitude of the drive laser on each site, and the laser detuning $\Delta \equiv \omega_{\rm L}-\omega_{\rm c}$.   The presence of hopping
changes these equations from their single-cavity counterpart.   We look for a 
solution to these equations that does not break the translational invariance of the Hamiltonian, i.e.~$\bar{a}_j = \bar{a}$ and $\bar{b}_j = \bar{b}$; the equations then
reduce to the single-site case with a shifted detuning $\Delta \rightarrow \Delta + 2 J$.  We assume that the drive is small enough so the equations do not exhibit
bistability \cite{Dorsel1983, Fabre1994, Weis2010, Florianreview}, and hence yield a single
solution for $\bar{a}, \bar{b}$; without loss of generality, we take $\bar{a}$ to be real and positive.

With these classical solutions in hand, we next make standard canonical displacement transformations of photon and phonon operators.  
We simply shift each $\hat{b}_j$ by its mean value, and define 
$\hat{d}_j = \hat{a}_j - \bar{a}$ (again, in a rotating frame with respect to the control laser frequency).
The resulting Hamiltonian has an optomechanical interaction 
enhanced by a factor of $\bar{a}$.  Assuming this factor to be large, we make the standard linearization, dropping interaction terms that are independent of $\bar{a}$.
We are assuming the standard regime where the single-photon optomechanical coupling $g$ is so weak that it can only play a role when enhanced by the amplitude induced
by the strong control laser.
The resulting coherent 
(i.e.~dissipation-free) Hamiltonian has the form:
\begin{eqnarray}\label{eqn:Hamiltonian}
	\hat{H}_{\rm coh} &=&\sum_j \left[-\Delta' \hat{d}^\dag_j \hat{d}_j 
	+ \Omegam\hat{b}^\dag_j \hat{b}_j 
	+ G(\hat{d}^\dag_j+ \hat{d}_j) ( \hat{b}^\dag_j+\hat{b}_j ) \right]
	 \nonumber \\
	&&
	-J\sum_{\langle i,j\rangle}\hat{d}^\dag_i \hat{d}_j,
\end{eqnarray}
where $G=\bar{a}_j g$ is the many-photon optomechanical coupling.  We stress that   
$G$ can be tuned by changing the control laser intensity.  The effective laser detuning in the displaced frame $\Delta' = \Delta - g (\bar{b} + \bar{b}^*)$; 
we just denote this $\Delta$ in what follows.  Note that  our primary interest in this work is on the lattice analogue of OMIT and not on strong-coupling
effects associated with the many-photon optomechanical coupling $G$.  As such, 
we will focus on regimes where $G \lesssim \kappa_0$, $G \ll  \Omegam$.

Given its translational symmetry, the Hamiltonian has a simpler structure in momentum space.  Setting the lattice constant
$a = 1$, we have: 
\begin{eqnarray}
	\hat{H}_{\rm coh}	&=&
		\int^{\pi}_{-\pi} dk 
		\Big(
			\omega_k\hat{d}_k^\dag\hat{d}_k
			+\Omegam \hat{b}_k^\dag\hat{b}_k 
		\nonumber \\
	&&
	 +G(\hat{d}_k^\dag\hat{b}_k+\hat{d}_k^\dag\hat{b}_{-k}^\dag+h.c.).
	 \label{eq:Hk}
\end{eqnarray}
Here, $\hat{d}_k$ and $\hat{b}_k$ are the momentum-space photon and mechanical 
fields ($-\pi \leq k < \pi$), and have a delta-function normalization 
(e.g.~$[\hat{d}_k, \hat{d}^\dag_{k'}] = \delta(k-k')$).  We use
\begin{equation}\label{eq:barephotonband}
	\omega_k=-\Delta-2J\cos{k}
\end{equation}
to denote the dispersion relation of the bare photon band (i.e.~in the absence of optomechanical interaction).
The corresponding bare phonon band has no dispersion, consistent with our neglect of phonon hopping.  
In the interaction picture, the position of the bare phonon band (described by Eq.~(\ref{eq:barephotonband})) relative 
to the flat photon band at $\omega = \Omegam$ can be controlled by the control laser detuning $\Delta$.  

Note that if one drops
terms in $\hat{H}_{\rm coh}$ which do not conserve excitation number, then the Hamiltonian for each invariant $k$ subspace is equivalent to an effective 
single-site optomechanical cavity in the same approximation, with an effective laser detuning $\Delta_\text{eff}=-\omega_k$.  In contrast, including the 
excitation number non-conserving 
terms ruins the correspondence to the single site problem, as now a given photon mode with momentum $k$ couples both to phonons with momentum $k$ and $-k$.

 
\subsection{Polariton bandstructure}
\label{subsec:BandStruture}

\begin{figure*}[htpb]
\begin{center}
\includegraphics[scale=0.35]{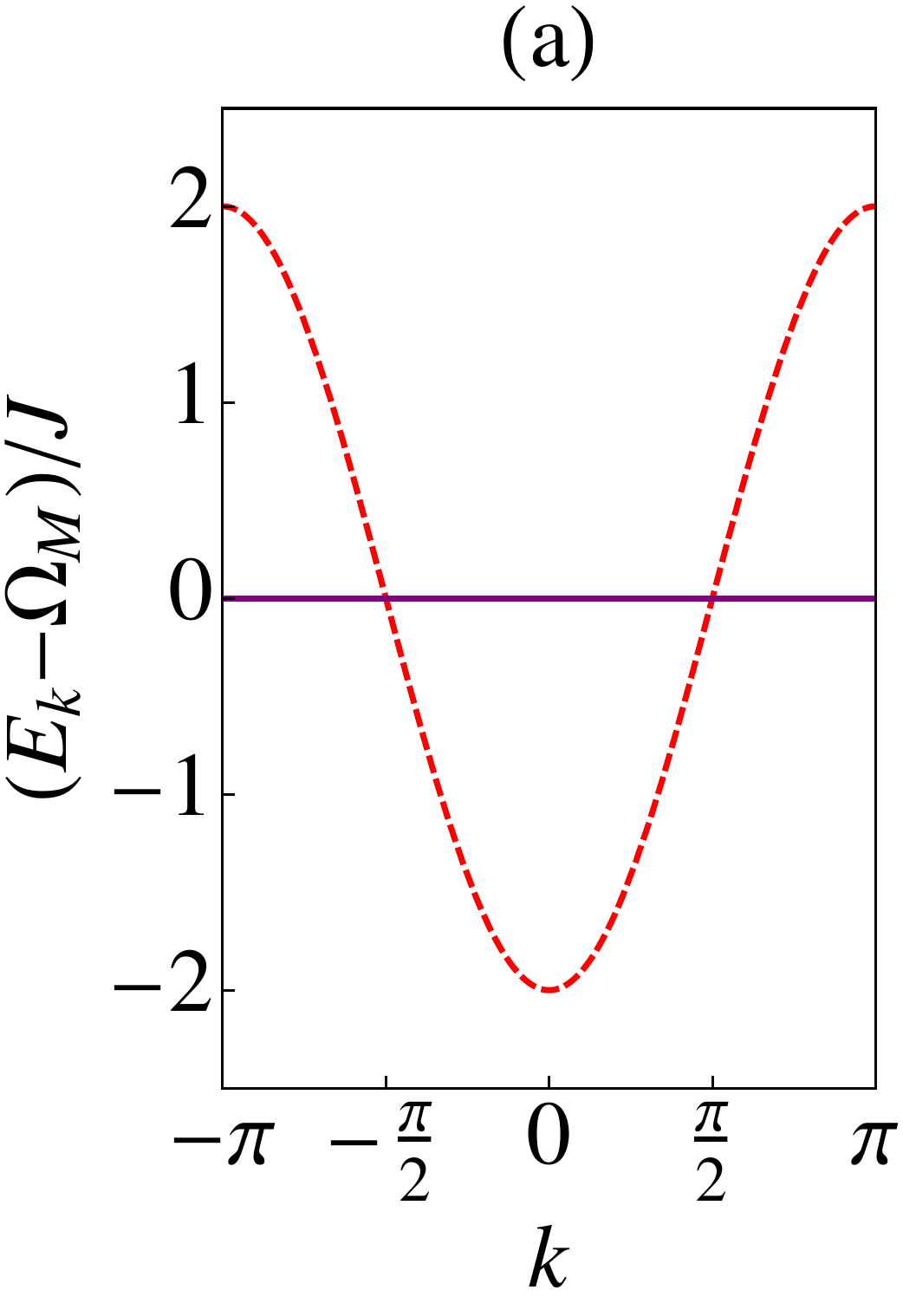}
\includegraphics[scale=0.35]{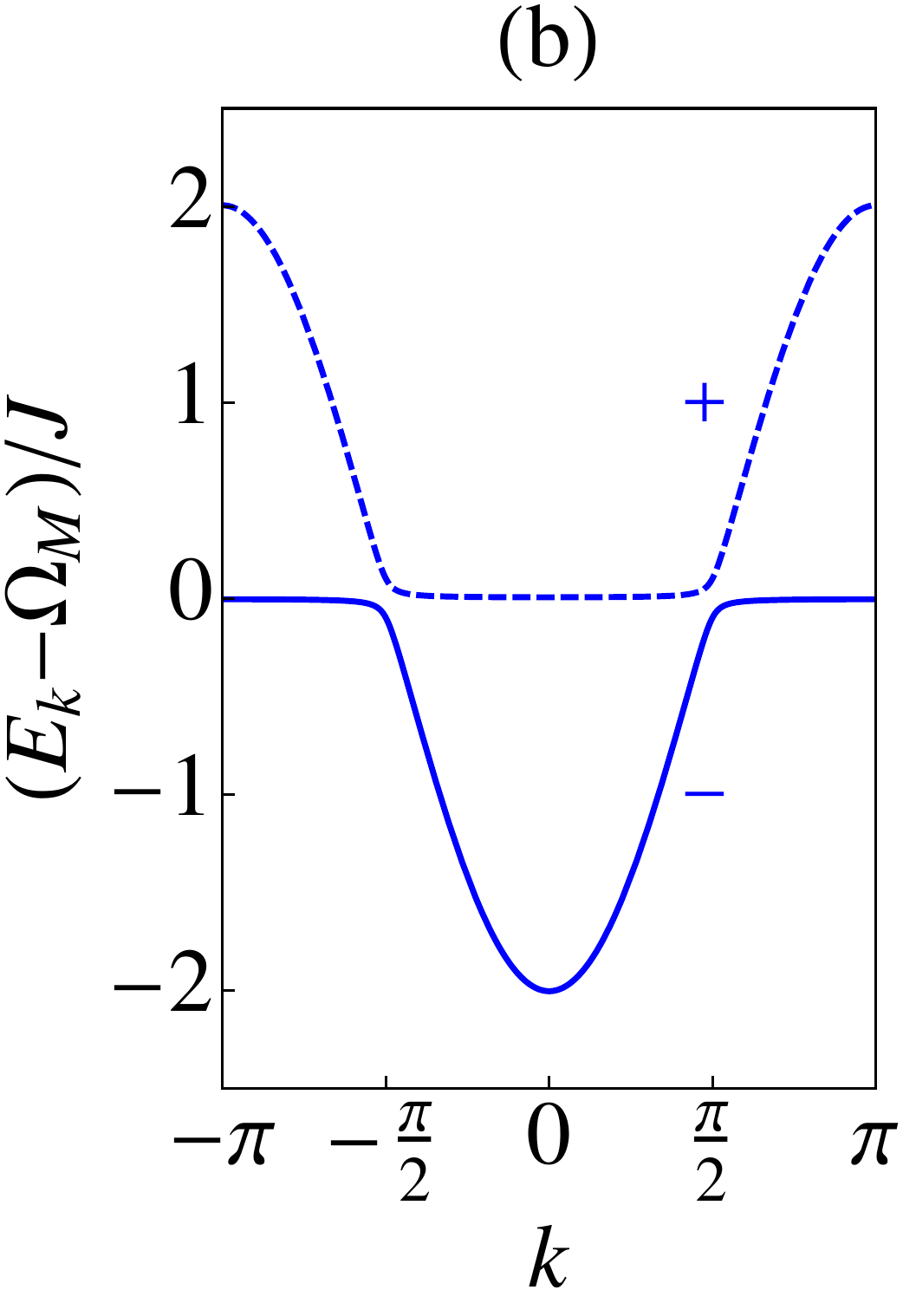}
\includegraphics[scale=0.35]{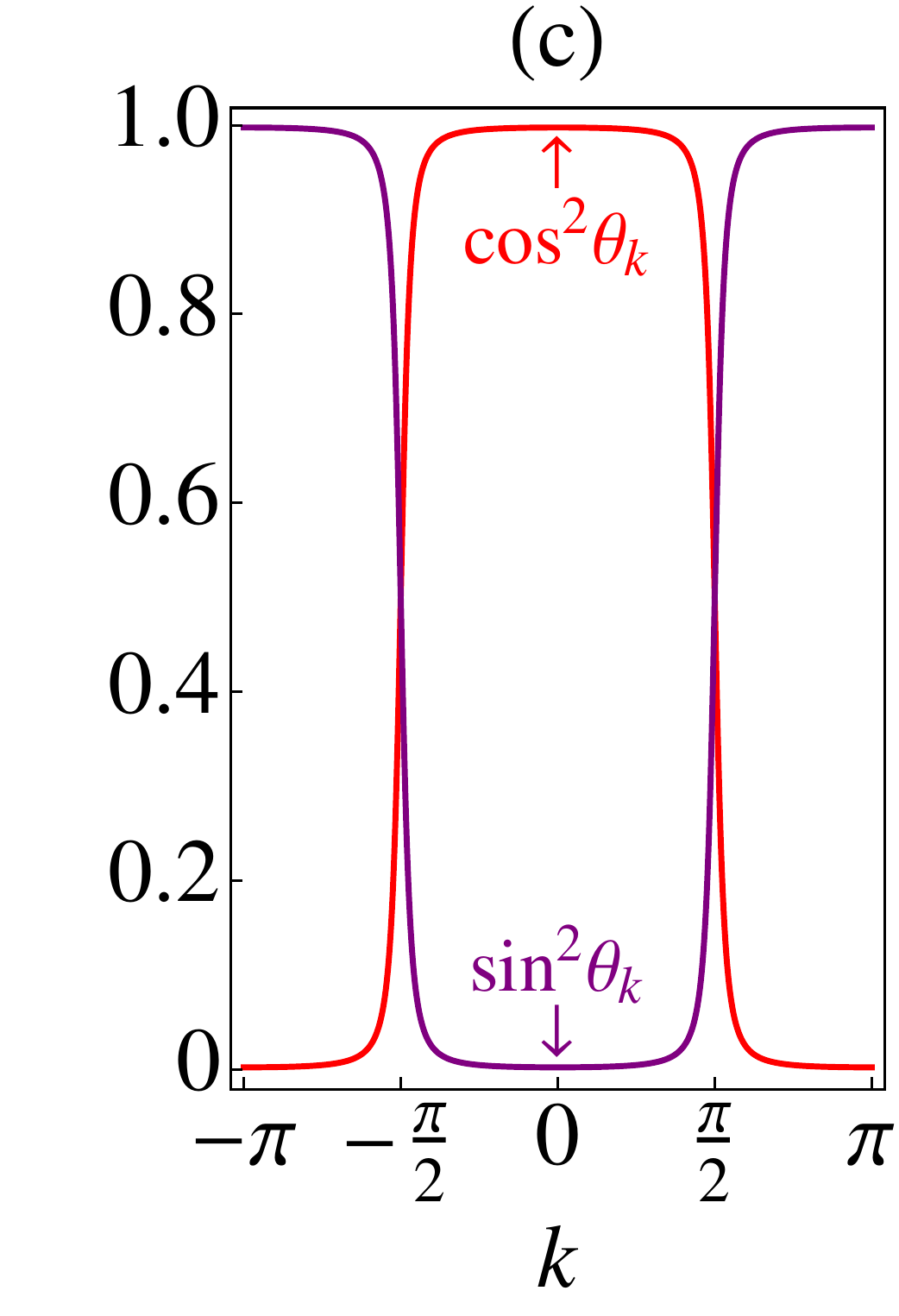}
\caption{(color online) (a): Band structure of the bare photons (red dashed curve) 
and mechanical modes (purple solid curve) for red sideband detuning 
$\Delta = -\Omegam$ and $G \rightarrow 0$.    
(b): Band structure of the polaritons, same parameters in (a) but now $G=0.1J$.
The upper (dashed) curve represents the $+$ polariton band 
and the lower (solid) curve represents the $-$ polariton band. 
(c) Variation of the factors $\sin^2 \theta_k$, $\cos^2 \theta_k$ which describe the photon and phonon parts of the polariton wavefunctions,
c.f.~Eqs.~(\ref{eq:polaritontransformation}). 
}
\label{fig:RSBbands}
\end{center}
\end{figure*}

The coherent Hamiltonian $\hat{H}_{\rm coh}$ in Eq.~(\ref{eq:Hk}) is readily diagonalized in terms of eigenstates that are mixtures of phonons and photons; we term these excitations polaritons.  

\subsubsection{Parametric instabilities}

An immediate issue is that, similar to a single-cavity optomechanical system, the excitation non-conserving ``parametric-amplifier" terms in Eq.~(\ref{eq:Hk}) can give rise to 
instabilities (see also Ref.~\onlinecite{Schmidt2013}).  Note that in the interaction picture, the bottom of the bare photon band is at an energy $\omega_{k=0} = -\Delta - 2J$; it can be negative even when $\Delta < 0$ 
(i.e.~the control laser is detuned to the red of the middle of the bare photon band).  The existence of negative energy photon states can make the parametric-amplifier terms in Eq.~(\ref{eq:Hk}) resonant, making the system especially susceptible to instability.  In particular, for a detuning $-\Delta \leq 2J - \Omegam$, we will have such a resonance for some value of $k$.  Including dissipation, a parametric instability will occur for such detunings when the cooperativity $4 C\equiv G^2/\gamma\kappa_0 >1$~\cite{Florianreview}.  Our focus in this work will be on regimes free of such instabilities.  As we are interested in weak optomechanical couplings $G \ll \Omegam $, we will thus restrict attention to laser detunings $\Delta$ that position the bare photon band to lie above 
$\omega = 0$ in the interaction picture.   In such cases, the parametric amplifier terms are off-resonant by a minimum amount $\Omegam$, and parametric instabilities cannot occur for optomechanical couplings $G \ll \Omegam$.

 \subsubsection{Rotating-wave approximation}

Given the choice of detunings and parameters discussed above, we will be in a regime where the effects of the particle non-conserving terms in Eq.~(\ref{eq:Hk}) are extremely weak.
As such, we can make a rotating-wave approximation (RWA) and drop them.  The resulting Hamiltonian is diagonalized as:
\begin{eqnarray}\label{eq:Hkspace}
	\hat{H}	&=& \int_{-\pi}^\pi dk \left[ \omega_k\hat{d}_k^\dag\hat{d}_k
	+ \Omegam \hat{b}_k^\dag\hat{b}_k+G(\hat{d}_k^\dag\hat{b}_k+h.c.)  \right] \nonumber\\
	&=& \sum_{\sigma = \pm}  \int_{-\pi}^\pi dk  \, E_{\sigma,k} \hat{\psi}_{\sigma,k}^\dag\hat{\psi}_{\sigma,k}.
\end{eqnarray}
Here, $\sigma=+,-$ denotes the two polaritons (normal-modes) at each momentum $k$ (energy $E_{\sigma,k}$, 
annihilation operator $\hat{\psi}_{\sigma,k}$).  One finds that the energies are
\begin{equation}
	\label{eq:polaritonenergy}
	E_{\pm,k}=\Omegam+\frac{\delta_k}{2}\pm\sqrt{G^2+\left(\frac{\delta_k}{2}\right)^2},
\end{equation}
where 
\begin{equation}\label{eq:delta_k}
	\delta_k	\equiv  \omega_k - \Omegam	
		=	\left(-\Delta-2J \cos{k}\right) -\Omegam
\end{equation}
is the energy detuning between the bare photon and phonon bands at wavevector $k$.  Similarly, the 
polariton mode operators are 
\bse
\label{eq:polaritontransformation}
\begin{eqnarray}
	\hat{\psi}_{+,k} & = &
		\cos{\theta_{k}} \cdot \hat{d}_k +  \sin{\theta_{k}} \cdot  \hat{b}_k,  \\
	\hat{\psi}_{-,k} & = &
		-\sin {\theta_{k}} \cdot \hat{d}_k +\cos{\theta_{k}} \cdot \hat{b}_k,
\end{eqnarray}
\ese
where the mixing angle $\theta_{k}$ is given by
\begin{eqnarray}
	\tan 2 \theta_{k} = 2 G / \delta_k.
	\label{eq:ThetaDefn}
\end{eqnarray}

Fig.~2 shows a typical polariton bandstructure (both with and without the optomechanical interaction) in the simple case where the bare photon band is bisected
by the bare mechanical band (i.e.~$\Delta = - \Omegam$).  While the RWA Hamiltonian in Eq.~(\ref{eq:Hkspace}) describes a simple two-band bosonic tight-binding model, it does not tell the complete story.  As we will see in what follows, the {\it dissipative nature} of our system will play a crucial role in determining the spectral and transport properties we are interested in (namely the fact that photonic and phononic damping rates are vastly different from one another). This makes our system markedly different from standard two-band models studied in the condensed matter literature.

\section{Cavity photon density of states and OMIT effect in an optomechanical array}
\label{sec:DOS}

Having discussed the simple properties of the dissipation-free, linearized optomechanical lattice, we
now turn to the more interesting and subtle features that arise when dissipation is included.  We first will examine the
local photon density of states (DOS) of the lattice, a quantity that can be directly measured via OMIT-style experiments where one looks at the
local reflection of an auxiliary probe laser; we explicitly calculate this reflection coefficient in Sec.~\ref{subsec:OMITreflection}.
One might naively expect that dissipation would simply broaden the density of states features present in the coherent system.  Similar to the
single site case, this is not the case:  the fact that the mechanical damping rate $\gamma$ is vastly different in magnitude than the the photonic
damping rate $\kappa_0$ leads to features and physics that go well beyond a simple energy broadening.  


\begin{figure*}[htbp]
\begin{center}
\includegraphics[scale=0.35]{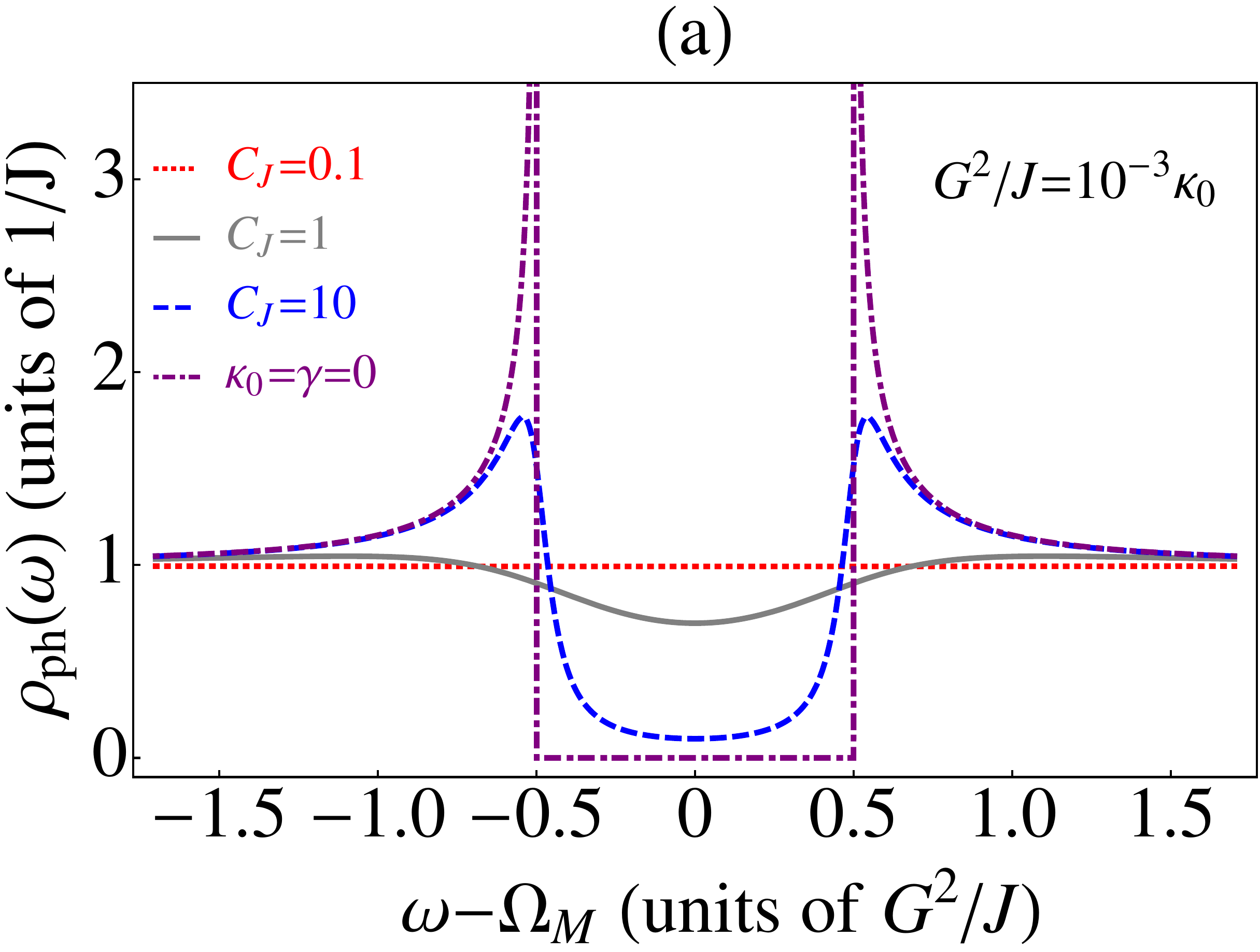}
\includegraphics[scale=0.36]{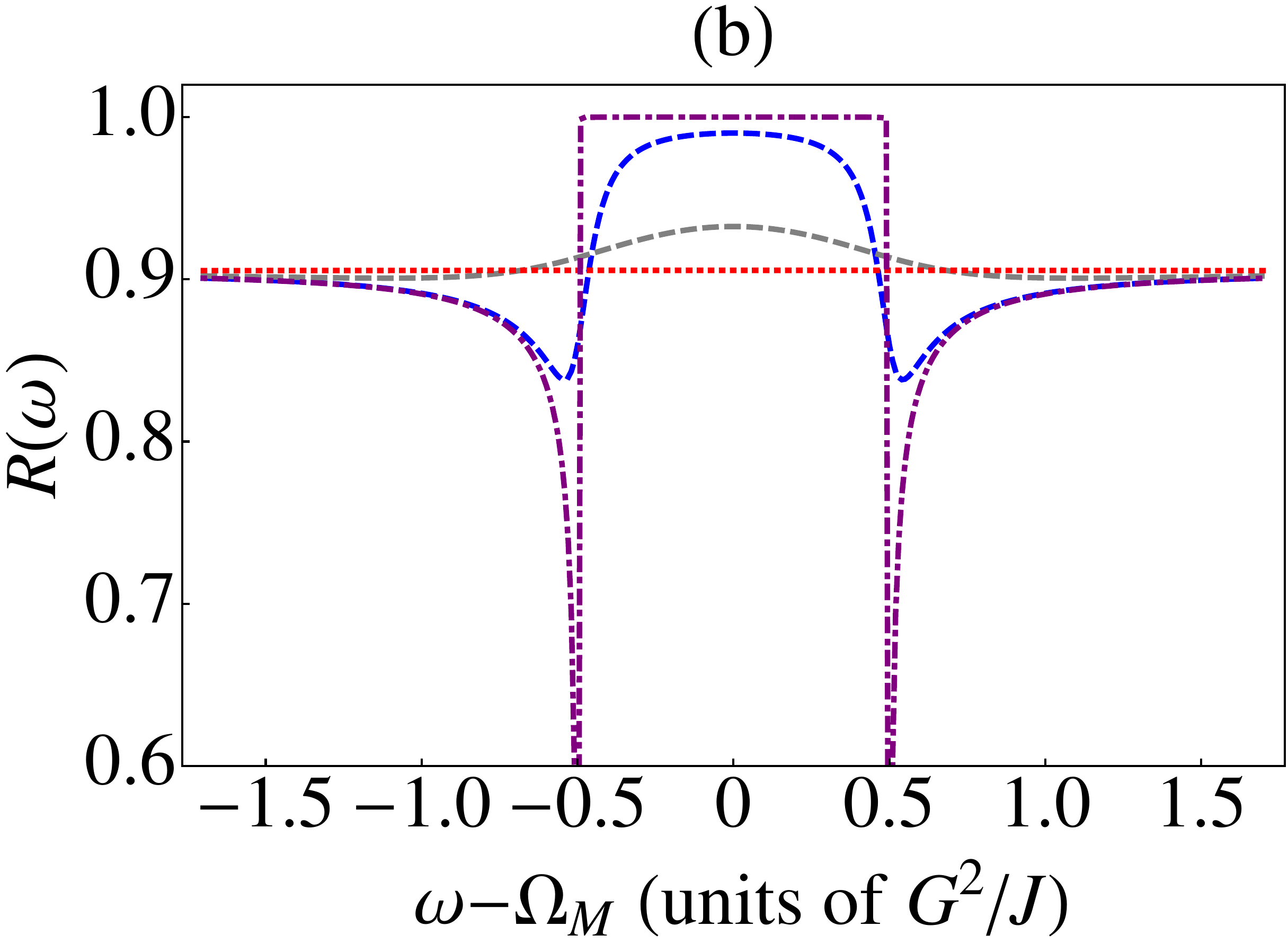}
\caption{(color online) 
(a): The local cavity photon DOS in the optomechanical array  
(c.f.~Eq.(\ref{eqn:DOS})) for frequencies near the energy gap in the coherent bandstructure;
the detuning $ -\Delta=\Omegam$ 
and optomechanical coupling $G=0.01J$ for all the curves (see Fig.~2 for a representative coherent bandstructure). $\kappa_0=0.1J$ 
for all the curves except the purple curve (dot-dashed) which corresponds 
to $\kappa_0=\gamma=0$. The different cooperativities 
in the figure are achieved by tuning the mechanical decay rate $\gamma$.
One sees that if the effective cooperativity $C_J$ is larger than $\sim 1$, the gap
in the coherent-system density of states persist, even if its magnitude is smaller 
than $\kappa_0$.
(b): The local reflection probability $R[\omega] = |r[\omega]|^2$ (Eq.(\ref{eq:reflection})) at 
frequencies near $\Omegam$ (center of the coherent-system gap) for a single weak 
probe on site $1$ with $\kappa_\text{ex,1}=0.1J$. Curves correspond to legend in 
panel (a). $R[\omega]$ directly mirrors the behaviour the DOS in (a). 
 }
\label{fig:RSBDOS}
\end{center}
\end{figure*}

\subsection{Calculation of photon density of states}
\label{subsec:DOScalc}

The local photon DOS on site $j$ of the lattice, $\rho_c(\omega,j)$ measures how hard it is to add a photon of energy
$\omega$ to site $j$.  It is formally defined as
\begin{equation}
	\rho_c(\omega, j)=-2 {\rm Im} D^R_{\rm c}(\omega; j,j), 
\end{equation}
where $D^R_c(\omega;j,j')$ is the retarded real-space Green function (GF) of cavity photons, defined
in the standard manner:
\bse
\begin{eqnarray}
	D^R_{c}(j,t;j',t') & = & -i\theta(t-t')\langle [\hat{d}_j(t), \hat{d}_{j'}^\dag(t')]\rangle, 
		\label{eq:GRDefn}\\
	D^R_{\rm c}(\omega; j,j') & = & \int^{\infty}_{-\infty} dt \ D^R_{c}(j,t';j',0 )e^{i \omega t}.
\end{eqnarray}
\ese
As we are dealing with a translationally invariant system, $\rho_c(\omega,j)$ will be independent of $j$.

The retarded GF of the cavity photons in $k$ space is easily obtained 
from the Heisenberg-Langevin equations of motion corresponding to our RWA Hamiltonian Eq.~(\ref{eq:Hkspace}) 
and a standard input-output treatment of the dissipative baths (see Appendix \ref{appendix:EOM})
\begin{equation}\label{eq:GFkspace}
	D^R_{\rm c}(\omega, k)=\frac{1}{\omega+\Delta+ 2J \cos{k}
	+ i\frac{\kappa_\text{0}}{2}-\frac{G^2}{\omega-\Omegam+i\frac{\gamma }{2}}}.
\end{equation}
For each $k$ mode, this  has exactly the same form as the photon GF for a single optomechanical cavity
(also treated within the RWA) with an effective $k$-dependent laser detuning.

Transforming to real space yields
\begin{eqnarray}\label{eq:photonGF}
	D^R_{c}(\omega; N,1)&=&\frac{1}{2\pi}\int^{\pi}_{-\pi}  
	dk\ e^{i k(N-1)} D^R_{c}(\omega, k)\nonumber\\
	&=&\frac{e^{iq[\omega](N-1)}}{2i J \sin{q[\omega]}}.
\end{eqnarray}
Here $q[\omega]$ is complex and satisfies
\begin{equation}\label{eq:k}
	\omega+\Delta+ i\frac{\kappa_0}{2}+ 2J \cos{q[\omega]}
	-\frac{G^2}{\omega-\Omegam+i\frac{\gamma}{2}}=0.
\end{equation}
As usual, the relevant solution here is the one with $\textrm{Im } q > 0$ (which thus yields behaviour consistent with causality).  It is useful
to decompose $q[\omega]$ into its real and imaginary parts, hence we write:
\begin{eqnarray}\label{eq:Z}
	q[\omega]\equiv\tilde{k}[\omega]+i\alpha[\omega]/2.
\end{eqnarray}
$\tilde{k}[\omega]$ represents the effective propagation wavevector at frequency $\omega$,
while $\alpha[\omega]$ describes the spatial attenuation of a wave at this frequency.

Using Eq.~(\ref{eq:photonGF}), we thus find that the local photon DOS is given by
\begin{eqnarray}\label{eqn:DOS}
	\rho_c(\omega,j) & = &
	\text{Re} \frac{2}{\sqrt{4J^2-(\omega+\Delta +i \frac{\kappa_0}{2}
	-\frac{G^2}{\omega-\Omegam +i \frac{\gamma }{2}})^2}}.\nonumber\\
\end{eqnarray}

Without dissipation (i.e.~$\kappa_0, \gamma \rightarrow 0$), 
the local density of states will exhibit the expected features corresponding to the coherent bandstructure:  there will be a gap for frequencies between
the two polariton bands, and there will be van Hove singularities at frequencies corresponding to the edges of each band.  One might expect that dissipation simply smears out
these features:  we now show explicitly that this is not the case.  In particular, we will see that even if the photon dissipation $\kappa_0$ is larger than the gap in the coherent bandstructure, 
a gap in the density of states can survive.

\subsection{Resilience of the gap against photonic dissipation}
\label{subsec:DOSsymmetric}

We consider for simplicity the symmetric situation where the laser detuning is $\Delta = - \Omegam$, which results in the
bare photon band being bisected by the bare mechanical band (as depicted in Fig.~\ref{fig:RSBbands}).
As discussed, we focus on the realistic situation
 where the optomechanical coupling is weak:  $G < \kappa_0$.  We also consider the photon hopping to be much larger than the internal photonic loss rate, but keep it small enough that our RWA treatment remains valid for our choice of detuning:  $\kappa_0 < J \ll \Omegam$.  

Recall first what happens in this regime in the absence of dissipation.  The optomechanical interaction leads to the formation
of two polariton bands, separated by an energy gap.  
The gap is centred on $\omega = \Omegam$ in the interaction picture, and has a width $\simeq G^2 / J$ 
(see Fig.~\ref{fig:RSBDOS}a, dashed-dotted curve).   
The polariton states with energies at the edges of this gap region (i.e. top of $-$ band, bottom of $+$ band) are almost entirely phonon-like:
the photon component of these polaritons is smaller by a factor $\sim G/J$ compared to the phonon component, c.f.~Eq.~(\ref{eq:polaritontransformation})
and  Figs.~\ref{fig:RSBbands}c,d.

Introducing dissipation, we would expect that if the photon damping rate $\kappa_0$ is larger than the size of the coherent-system gap $G^2 / J$, the gap will be smeared away.
Fig.~\ref{fig:RSBDOS} shows that this is not necessarily true: the gap can persist in the photon DOS even for relatively large values of $\kappa_0$.  

While at first glance this behaviour is surprising, a simple explanation is possible.  In the presence of dissipation, the 
phonon-like polariton modes with energies at the edges of the gap will have a total damping rate given by:
\bse
\begin{eqnarray}
	\label{eq:DampingGapEdge}
	\Gamma_{\rm gap-edge} \simeq \gamma + \Gamma_{\rm opt} ,\\
	\Gamma_{\rm opt} \sim   \left( G / J \right)^2 \kappa_0,
\end{eqnarray}
\ese
$\Gamma_{\rm opt}$ is an effective optical damping of these phonon-like modes:  it is given simply by the intrinsic cavity damping rate $\kappa_0$, times the small but non-zero
photon weight of these polariton modes, $\cos^2 \theta_k\sim (G/J)^2$.
 We see that these modes see only weakly the photon loss rate $\kappa_0$.
The approximate condition that now ensures that dissipation does not smear out the gap is that the energy-broadening of a gap-edge polariton $\Gamma_{\rm gap-edge}$
needs to be much smaller than the size of the gap in the coherent system, $G^2 / J$.  This can be recast as the conditions:
\begin{equation}
	\kappa_0 \ll J, \,\,\,\,\, C_J \gg 1
\end{equation}
where we have introduced the effective lattice cooperativity 
\begin{eqnarray}
	\label{eq:LCRSB}
	C_{J} \equiv  G^2/J\gamma.
\end{eqnarray}
We thus see that the gap in the photon density of states can remain pronounced even though it is much smaller than $\kappa_0$.  

We stress that the structure of the dissipation in our system plays a key role in obtaining this result (namely that there are two very different damping rates $\kappa_0$ and $\gamma$, and that the dissipation is not diagonal in the eigenbasis of the coherent Hamiltonian).  The dissipative physics in our system cannot be accurately described by introducing a simple homogeneous energy broadening into our system.   
If one has $\kappa = \gamma$, then there is no gap feature left, one gets essentially the red (dotted) curves in Fig. 3a, b.


\subsection{Comparison with single-cavity OMIT}

The fact that the gap in the photon density of states persists even when it is much smaller than the photon loss rate $\kappa_0$ is analogous to the situation in standard, single-cavity OMIT \cite{Agarwal2010,Weis2010}. We begin by summarizing the essential features of this effect, which involves driving a single-cavity optomechanical cavity with a strong control laser
near the red mechanical sideband $\Delta = -\Omegam$.  One obtains a standard beam-splitter interaction which coherently converts photons to phonons and vice-versa. 
 In the absence of dissipation, the optomechanical cavity would thus have two normal modes (polaritons) split by an energy $2 G$.  As a result, in the rotating frame, the cavity density of states would be zero for frequencies between $\omega-\Omega_M = -G$ and $\omega-\Omega_M = G$.  Introducing dissipation, one would naively think that this density of states suppression would be completely lost if $\kappa_0 \gg G$.  Instead, one finds that there is a density-of-states suppression in this regime at $\omega = 0$ as long as the standard optomechanical cooperativity $C = 4 G^2 / \kappa_0 \gamma$ is larger than one.  Further, the width of the DOS suppression in this regime is set by the effective mechanical linewidth (which includes an optical damping contribution); this width is much smaller than $G$.  This is the standard single-cavity OMIT effect; it can be interpreted as arising from interference between different pathways for adding a photon to the cavity, and can be measured by looking at the transmission or reflection of an auxiliary probe laser.

Returning now to our lattice system, we find a similar situation to the single cell case described above (again for $G \ll \kappa_0$), but with some important differences.  First, it is now the lattice cooperativity $C_J$ which plays a crucial role in determining whether the gap persists.  Second, the spectral width of the DOS suppression is about the same as in the coherent system, $\sim G^2 / J$; this is in contrast to single-site OMIT, where the width of the DOS suppression is much smaller than the ``gap" in the coherent system $2 G$.


\subsection{Measuring OMIT  in an optomechanical array}
\label{subsec:OMITreflection}

Similar to the situation with single-cavity OMIT, one can measure the DOS suppression described above by applying an additional weak probe laser with frequency $\omega_p$ 
to a site of the array, and measuring its reflection as $\omega_p$ is varied.  To model an OMIT-style experiment, we assume that an additional waveguide is coupled to a site of the lattice
(without loss of generality, we take this to be the site $j=1$).  The coupling to this waveguide is again treated using 
standard input-output theory \cite{Clerkreview}; it increases the damping rate on site $1$ from $\kappa_0$ to $\kappa_0 + \kappa_{1,{\rm ex}} \equiv \kappa_1$, 
where $\kappa_{1,{\rm ex}}$ is the coupling rate to the probe waveguide.  Site $1$ is now driven by an additional input field 
$\hat{d}_{1,{\rm p},{\rm in}}(t)$ describing the probe waveguide.  This field has a non-zero average value describing 
the applied probe tone:
\begin{equation}
	\langle\hat{d}_\text{1,p,in}(t)\rangle=\bar{d}_\text{1,p,in}e^{-i\omega_p t}.
\end{equation}

Input-output theory now tells us the corresponding average output field emanating back into the probe waveguide:
\begin{equation}
	\hat{d}_\text{1,p,out}=\hat{d}_\text{1,p,in}+\sqrt{\kappa_\text{ex,1} }\hat{d}_{1},
\end{equation}
The average value of this output field will give us the desired reflection coefficient of the probe field.  As our system is linear,
this quantity can be related directly to the retarded photon Green function (e.g.~by solving the averaged Heisenberg equations of motion, or
by using the Kubo formula).  One finds:
\begin{equation}\label{eq:reflection}
	r[\omega_p] = 
		\langle \hat{d}_\text{1,p,out} \rangle/\langle \hat{d}_\text{1,p,in} 
	\rangle=1-i \kappa_\text{ex,1}  \tilde{D}^R_{\rm c}(\omega_p; 1,1).
\end{equation}
The GF $\tilde{D}^R_{\rm c}(\omega_p; 1,1)$ is defined as per Eq.~(\ref{eq:GRDefn}), except that it now includes the effects of the coupling to the probe waveguide (i.e. the additional
photon damping on site $j=1$).  Note that this coupling breaks the translational invariance of the model.

To calculate the retarded GF in the presence of the probe waveguide, we need to include the additional photon damping due to the probe waveguide.  This is equivalent to
introducing an extra non-Hermitian potential for photons on site $1$
\begin{equation}\label{eq:imaginarypotential1}
	\hat{V}=-i\frac{\kappa_\text{ex,1}}{2}\hat{d}^\dag_1\hat{d}_1,
\end{equation}
Solving the Dyson equation corresponding to this extra potential allows us to calculate the ``dressed" GF $\tilde{D}^R$ (see Appendix \ref{appendix:GFCalc}).  We find:
\begin{equation}\label{eq:GFlocaldressed}
	\tilde{D}^R_{c}(\omega; 1,1)=\frac{1}{2i J \sin{q[\omega]}+i\frac{\kappa_\text{ex,1}}{2}}.
\end{equation}
where the complex function $q[\omega]$ is again defined via Eq.~(\ref{eq:k}).

For the case of a weak probe waveguide coupling (i.e.~$\kappa_\text{ex,1}\ll J$), the ``dressed" GF 
$\tilde{D}^R_{c}$ is similar to the intrinsic GF $D^R_c$ of the probe-free system.  Thus,
as expected, the reflection coefficient as a function of probe frequency $r[\omega_p]$ directly
reflects the structure in the density of states discussed earlier.
This is demonstrated by comparing 
Fig.~\ref{fig:RSBDOS}b and Fig.~\ref{fig:RSBDOS}c.
On a heuristic level, for a weak-coupling, the only way for a probe photon
to {\it not} be reflected is if there is a state in the cavity for it to enter.  If it enters such a state, the much stronger coupling to the rest of the lattice means there is essentially
no chance that the photon will be reflected back into the coupling waveguide.  Hence, for weak coupling, large reflection is associated with small photon density-of-states.

\section{Transport of probe laser photons in the presence of optomechanical coupling}
\label{sec:Transport}

In the previous section, we examined how even a weak many-photon optomechanical interaction could lead to a suppression
of the photon DOS, and how this could manifest itself in the local reflection of a probe laser beam.  In this section, we now address
the more general and rich question of the transport of photons injected by a probe laser beam:  how does the probe light injected at site $1$ propagate through the lattice to site $N+1$?  

We will again focus exclusively on the experimentally-relevant parameter regime where the optomechanical coupling $G$ is small:  $G \ll J, \Omegam$.  Also, to avoid any possibility of parametric instability (and hence again justify the use of the RWA Hamiltonian in Eq.~(\ref{eq:Hkspace})), we will focus on control laser detunings $\Delta < - 2 J$; for this regime, the bare photon band (as given by Eq~(\ref{eq:barephotonband})) always has positive energy in our interaction picture.

\subsection{Calculation of transmission amplitude}

For a probe-light transmission experiment, we imagine coupling auxiliary waveguides to sites $1$ and $N+1$ of our lattice,
as shown in Fig.\ref{fig:setup}b.  These additional waveguide couplings are treated analogously to what was done in Sec.~\ref{subsec:OMITreflection}; 
they increase the photon damping rate on sites $1$ and $N+1$ to 
$\kappa_0 +\kappa_\text{ex,1} \equiv \kappa_1$ and $\kappa_0 + \kappa_{N+1,{\rm ex}} \equiv \kappa_{N+1}$ respectively.  
We assume that a weak probe tone is incident
on site $1$ at frequency $\omega_p$, and then ask what the induced amplitude of the outgoing light is leaving site $N+1$ through the probe waveguide.  Using standard input-output
relations combined with linear response (analogously to Sec.~\ref{subsec:OMITreflection}), we obtain the transmission coefficient:
 \begin{eqnarray}\label{eq:transmission}
	t[N,\omega_p]	&=&	
		\frac{ \langle \hat{d}_\text{N+1,out}(t) \rangle}{  \langle \hat{d}_\text{1,in}(t) \rangle}
		=	\frac{  \langle \hat{d}_\text{N+1,out}(t) \rangle }{  \bar{d}_\text{1,in} e^{-i \omega_p t} }	
		\nonumber\\
	&=&-i\sqrt{\kappa_\text{ex,1}\kappa_\text{ex,N+1}}  \tilde{D}^R_{\rm c}(\omega_p; N+1,1).
	\label{eq:tNMain}
\end{eqnarray}
Here, $\tilde{D}^R_{\rm c}(\omega_p; N+1,1)$ is the retarded photon Green function calculated in the presence of the couplings
to the probe waveguides (which break translational invariance).

We can again calculate $\tilde{D}^R_{\rm c}(\omega_p; N+1,1)$ by introducing an imaginary potential for cavity photons on sites $1$ and $N+1$, 
i.e.~$\hat{V}=-i \sum_{i=1,N+1} \frac{\kappa_\text{ex,i}}{2}\hat{d}^\dag_i\hat{d}_i$,
and then solve the resulting Dyson equation for the full GF.  The GF is then found to be (see Appendix \ref{appendix:GFCalc} for details)
\begin{widetext}
	\begin{eqnarray}\label{eq:GFdressed}
		\tilde{D}^R_{c}(\omega;N+1,1)=
			\frac{8i J \sin{q[\omega]} \ e^{iq[\omega]N}}{ \kappa_\text{ex,1}\kappa_\text{N+1,ex} (e^{i 2N q[\omega]}-1)
			- 4 J \sin{q[\omega]}( \kappa_\text{ex,1}+\kappa_\text{N+1,ex})-(4 J \sin{q[\omega]})^2},
	\end{eqnarray} 
\end{widetext}
where the complex propagation wavevector $q[\omega] \equiv \tilde{k}[\omega] + i \alpha[\omega]/2$ is independent of the waveguide couplings, and is again defined via Eq.~(\ref{eq:k}).


For more intuition, note that the presence of two coupling waveguides makes our system
somewhat analogous to a Fabry-Perot interferometer, as these couplings can induce multiple local reflections within the lattice.  It is thus useful to re-write the transmission
obtained from Eqs.~(\ref{eq:tNMain}) and  (\ref{eq:GFdressed})  in the standard form:
\begin{equation}
	t[N, \omega_p]=\frac{\tlt_1[\omega_p] \tlt_{N+1}[\omega_p] e^{ i q[\omega_p] N}  }{1-\tr_1[\omega_p]  \tr_{N+1}[\omega_p]e^{2iq[\omega_p] N}}.
	\label{eq:FPtransmission}
\end{equation}
$\tlt_i[\omega_p]$ $(i=1, N+1)$ describe the transmission of photons between the lattice and the 
two external waveguides, while $\tr_i[\omega_p]$ describes the back-reflection of photons in the lattice at sites $1$, $N+1$ due to the effective impedance mismatch caused by the external waveguide couplings.
One finds:
\begin{equation}
	\label{eq:r_and_t}
	 \tr_i  =  
	 	\frac{  \kappa_{\textrm{ex},i}}{\kappa_{\textrm{ex},i}+4J \sin{q[\omega]}}; 
	 \tlt_i=
 		\frac{\sqrt{8J\kappa_{\text{ex},i} \sin{q[\omega]}}} {\kappa_{\text{ex},i}+4J\sin{q[\omega]}}. 
 \end{equation}

Note that both the optomechanical interaction $G$ and the presence of internal dissipation in the lattice enter completely through the effective complex wavevector $q[\omega]$ (c.f.~Eq.~(\ref{eq:k})).  

\subsection{Differences from transport in atomic EIT systems}
\label{subsec:EITanalogy}

\begin{figure*}[t]
	\includegraphics[scale=0.36]{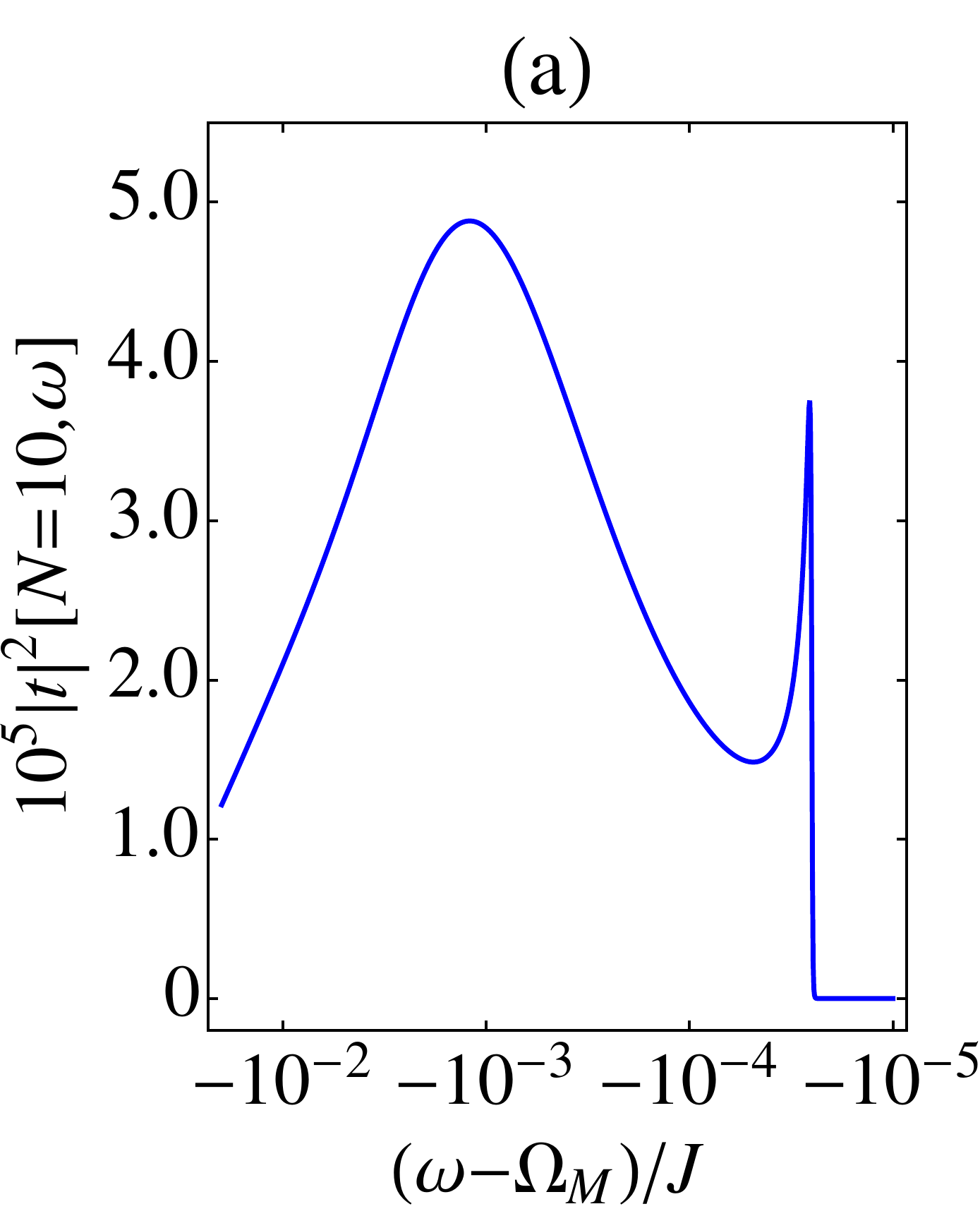}
	\includegraphics[scale=0.365]{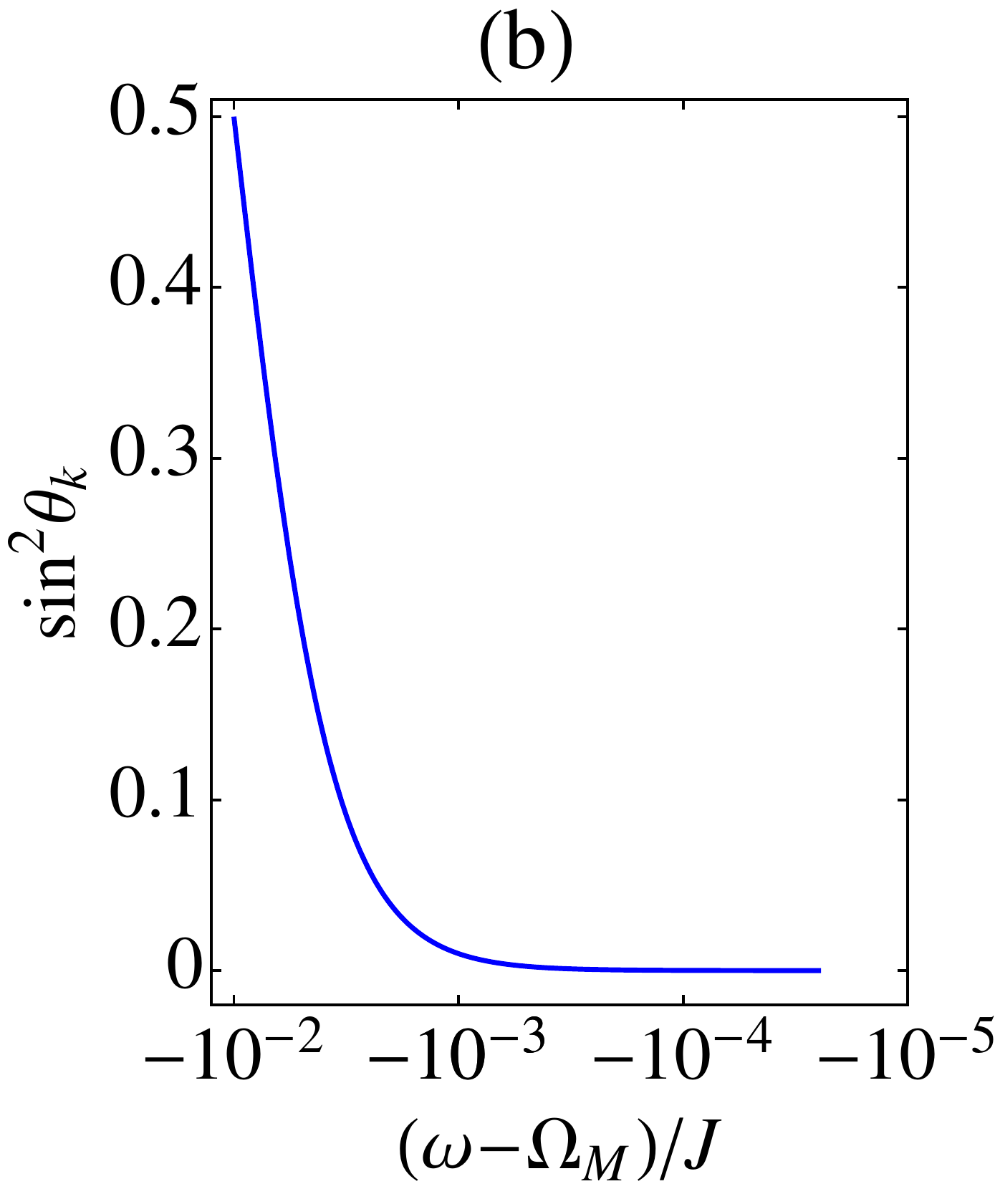}
	\includegraphics[scale=0.36]{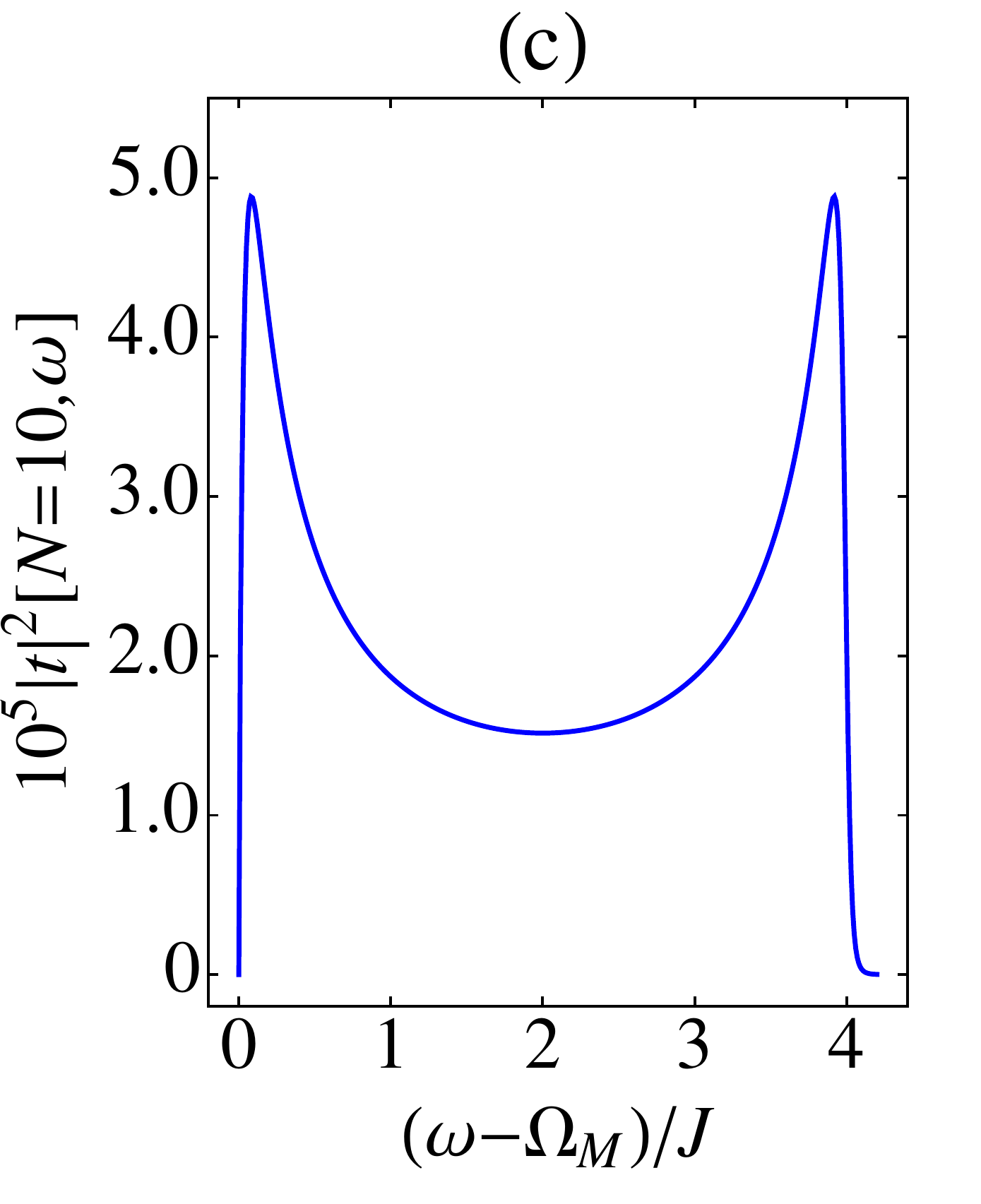}
	\caption{
		(color online) Transmission probability of probe light from sites 1 to 11
		as a function of probe frequency, for the case of a relatively weak coupling to probe waveguides
		($\kappa_\text{ex,1}=\kappa_\text{N,ex} = 0.01 J $), and for relatively large internal cavity loss
		($\kappa_0=0.1J$ ).  We have picked parameters such that the bare photon and phonon bands just touch at
		$k=0$ (i.e. $\delta_0=-\Delta-\Omegam-2J=0$), and taken $G =  0.1 \kappa_0, \gamma=10^{-6} \kappa_0$. Panel (a)
		shows the contribution from the lower, phonon-like polariton band (using a logarithmic $\omega$-axis), 
		panel (b) shows on the same scale the variation of the photonic component of the $-$ polariton as a function of frequency.
		Panel (c) shows the transmission probability from the photon-like upper polariton band.
		Surprisingly, both polariton bands can make comparable contributions to the photon transmission in this limit; this is due to the
		fact that the the optical damping of the lower polariton band $\Gamma_{\rm opt}$ is much larger than $\gamma$.
}
	\label{fig:WeakCouplingTrans}
\end{figure*}

Before proceeding with our analysis, it is worth pausing to emphasize a key difference between probe photon transport in our optomechanical lattice, and light propagation in standard atomic EIT systems.  Recall that atomic EIT involves three-level ``lambda"-system atoms, with two lower levels $|g1\rangle, |g2\rangle$ and an upper excited level $|e\rangle$.  In an atomic EIT medium, the physics can be best understood in terms of 
dark-state polaritons \cite{Fleischhauer2000,Fleischhauer2002}:  light propagates as a coherent superposition of a photonic excitation and an atomic excitation $|g1\rangle \rightarrow |g2 \rangle$, in such a way that destructive interference prevents the uppermost atomic level
$|e\rangle$ from being populated.  In the analogy to OMIT \cite{Agarwal2010,Weis2010}, the role of the $|g1\rangle \rightarrow |g2 \rangle$ excitation is played by a mechanical phonon excitation, while the role of the upper atomic excitation $|e \rangle$ is played by having a photon in the cavity.  Thus, the destructive interference that prevents excitation to $|e \rangle$ in 
atomic EIT now prevents a probe photon from entering the cavity in OMIT.

Turning to propagation physics, in our system probe photons can {\it only} be transported through the lattice by entering the cavity.  As such, this {\it cannot} be mapped onto an atomic EIT dark state polariton, as having a photon enter the cavity in our system is analogous to exciting an atom to the upper $|e \rangle$ state in the atomic system.  For this basic reason, the physics we describe here is not equivalent to light propagation in an atomic EIT medium \cite{Lukin1998}.  We note that the alternative system studied by Chang et al. \cite{Chang2011}, where optomechanical cavities are periodically side-coupled to a waveguide, is more closely analogous to this situation.  In that system, one can discuss 
probe photon propagation in terms of a dark-state polariton involving a probe photon in the waveguide, and a phonon in the mechanical mode; having a photon in the cavity is not necessary.

\subsection{Transmission for a weak probe waveguide - lattice coupling}
\label{subsec:WeakTransport}

For very weak probe couplings, the reflection amplitudes $\tr_1, \tr_{N+1}$ are negligible,
and the transmission is well approximated by the numerator of Eq.~(\ref{eq:FPtransmission}).  Equivalently, one can interpret the transmission as being a probe of the intrinsic photon GF $D^R_c(\omega;N)$, similar
to the local reflection coefficient in Eq.~(\ref{eq:reflection}) in this limit.  
However, in contrast to the reflection coefficient, one is now probing the intrinsic GF at non-zero propagation distance $N$.

The expectation would thus be that $t[N,\omega_p]$ is simply 
a weighted measure of the photon density of states, where the contribution of each polariton is weighted both by how photonic it is (i.e.~the appropriate factor of $\sin \theta_k$ or
$\cos \theta_k$, c.f.~Eq.~(\ref{eq:ThetaDefn})), and by a factor describing attenuation. Attenuation due to propagation is
described by the factor $\alpha[\omega]$ given in Eqs.~(\ref{eq:k}) and (\ref{eq:Z}).   In the absence of any optomechanical interaction, and for $\kappa_0 \ll J$, it is simply given by:
\begin{equation}
	\label{eq:AlphaNoG}
	\alpha[\omega] = \frac{ \kappa_0}{2 J \sin k[\omega] } = \frac{\kappa_0}{v_g[\omega]}
\end{equation}
where $v_g[\omega] = 2 J \sin k[\omega] $ is the group velocity of the bare photon band, and $k[\omega]$ is the photon wavevector corresponding to the band energy $\omega$. 

Based on the above, one might conclude that polaritons that are almost entirely phonon-like will make a negligible contribution to the transmission of probe photons.  This is incorrect, as is shown in Fig.~\ref{fig:WeakCouplingTrans}.  Here, we pick parameters such that the bare phonon and photon bands only touch at $k=0$.  As such, the lower polariton band is almost entirely phonon-like (except for a narrow range of $k$ vectors near $k=0$, i.e. near the bottom of the $-$ band).  The figure shows the transmission probability of probe light from site $1$ to $11$, in a regime where the internal cavity loss is relatively large.  We see that, surprisingly, the phonon-like lower polariton band makes an appreciable contribution to the transmission over its full width:  it yields maximum transmissions that are on par with
those associated with the photon-like upper polariton band.  

The explanation is straightforward:  even though most $-$ band polaritons are largely phonon-like, their dissipation (for these parameters) is dominated by photonic losses ($\kappa_0, \kappa_{\rm ex}$) associated with the small photon-part of their wavefunction.  As a result, they are effectively as well-coupled to the external probe waveguides as states in the more photon-like $+$ polariton band.  Focusing on $k$ vectors away from $k=0$, we see from Eq.~(\ref{eq:polaritontransformation}) that the photonic amplitude associated with a $-$ polariton scales as $\sin \theta_k \sim G / J \ll 1$.  As such, the ``optical damping" of such a mode will be  
\begin{equation}
	\Gamma_{\rm opt} \sim	\frac{G^2}{J^2} \kappa_{0}.
\end{equation}
If this induced optical damping dominates the intrinsic damping $\gamma$ of the phonon-like modes in the $-$ band, then these modes can be used for probe photon transport with a transmission on par with the purely photonic modes in the $+$ polariton band.

The transmission profile associated with the photon-like $+$ polariton band in Fig.~\ref{fig:WeakCouplingTrans} reflects the shape of the corresponding DOS, exhibiting 
van-Hove singularities at the band edges.  In contrast, the transmission profile of the phonon like $-$ polariton band is more asymmetric; this reflects the  variation of the mixing angle $\theta_k$ (c.f.~Eq.~(\ref{eq:ThetaDefn})) through the Brillouin zone.  


\subsection{Transport for strong waveguide coupling: interplay of Fabry-Perot interference and OMIT}

\begin{figure*}[htbp]
	\includegraphics[width= \textwidth]{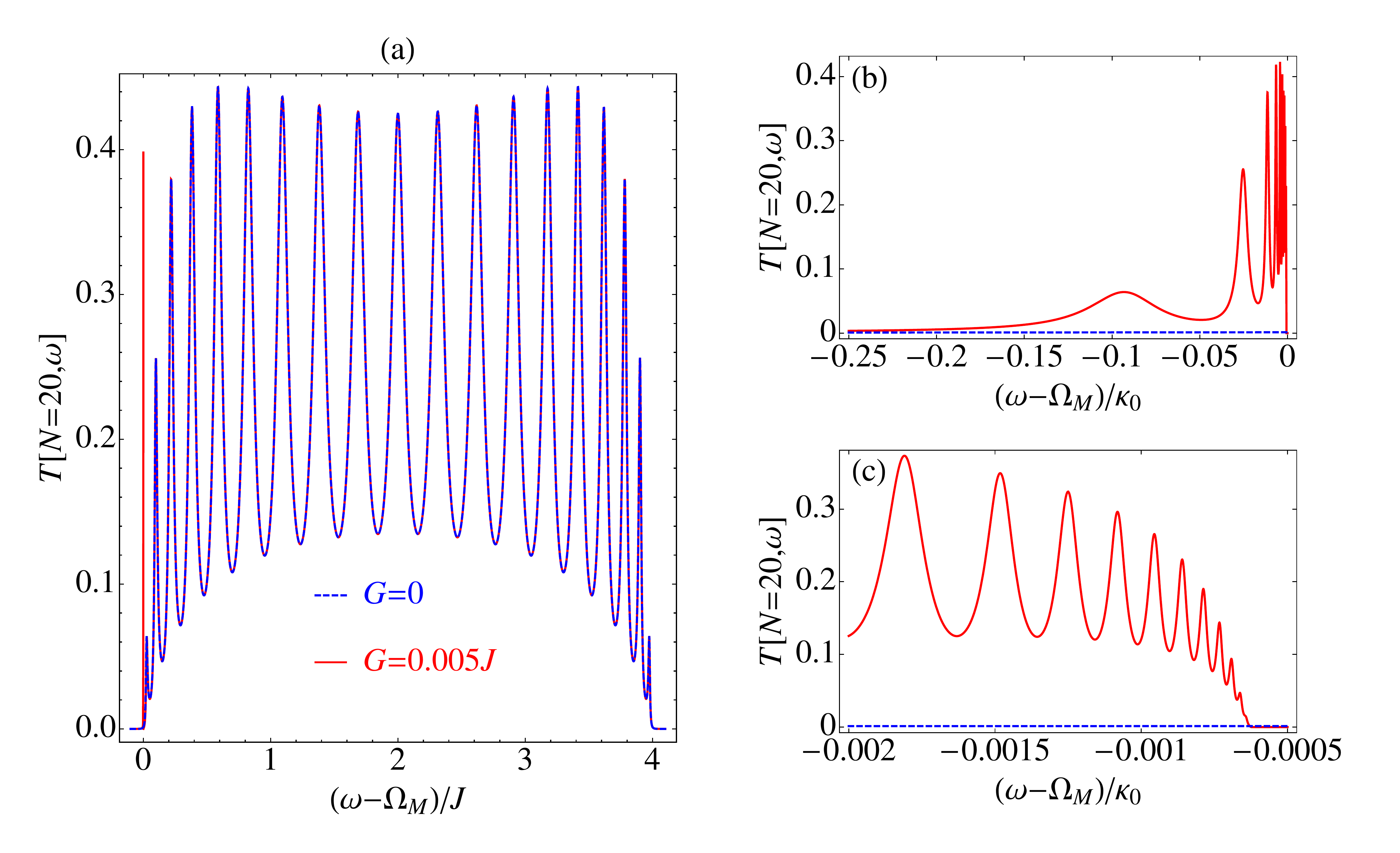}
	\caption{
		(color online) Transmission probability for probe light from sites $j=1$ 
		to $j=21$, as a function of the probe beam frequency $\omega$, in the regime of a strong coupling to 
		the external probe waveguides:  $\kappa_{\rm ex} = 5 J$.  We also take a control laser detuning yielding $\delta_0 = 0$ (c.f.~Eq.~(\ref{eq:delta_k})), implying
		that the bare photon band just touches the bare phonon band at $k=0$.  (a) Blue curve: transmission spectrum without
		optomechanical coupling, $G=0$, and with $\kappa_0  = 0.01 J$.  One sees well-developed Fabry-Perot style resonances
		associated with the bare photon band.  Red curve:  same, but now with $G = \kappa_0/2$ and $\gamma = 10^{-5} \kappa_0$.  The main effect of increasing the 
		optomechanical coupling is the emergence of an extremely sharp transmission feature near $\omega \sim \Omegam$, 
		associated with the phonon-like $-$ polariton band.  (b), (c)  Zoom in of the $-$ polariton band features in (a).  
		One again sees Fabry-Perot style resonances, leading to features in the transmission that are much narrower that
		$\kappa_0$.} 
	\label{fig:FPresonances}
\end{figure*}

We now consider probe photon transmission in the regime where the couplings to the probe waveguides are large, $\kappa_{\rm ex} > J$.
Unlike the weak-coupling regime considered in the previous subsection, transmission probabilities of order unity are now possible.  From Eqs.~(\ref{eq:FPtransmission})
and (\ref{eq:r_and_t}), we see that a new aspect
of the physics in this regime will be geometric, Fabry-Perot style resonances caused by constructive interference of photons in the lattice reflecting between sites $1$ and $N+1$.  Understanding how these geometric resonances are modified by the optomechanical coupling and the physics underlying OMIT will be a key goal of this subsection.
Note that the same interplay of OMIT physics and Fabry-Perot interference can happen in a {\it finite} lattice with probe coupling at the end sites $1$ and $N+1$, resulting in similar transmission properties, see Appendix \ref{appendix:Finite} for details.

We again focus on the simple case where bare photon and phonon bands only touch at $k=0$, and where $G \ll \Omegam, J$; for
such a system, the $-$ polariton band is mostly phonon-like, whereas the $+$ band is mostly photon-like.  Shown in Figs.~\ref{fig:FPresonances}(a)-(c) is an example of the transmission
spectrum in this regime, both with and without the optomechanical coupling $G$.  Without the coupling (Fig.~\ref{fig:FPresonances}a), one sees the expected pattern of Fabry-Perot resonances associated with the photon band; resonances occur when
\begin{equation}
	\label{eq:resonancek}
	k[\omega] =   \pm(\pi-\frac{m}{N}\pi) \equiv \pm k_m , \ \ m=0, 1, ..., N.
\end{equation}

Turning on the optomechanical interaction but keeping it weak ($0 < G < \kappa_0$), we see that the main features present for $G=0$ persist; these are now resonances associated with the $+$ polariton band, which is mostly photonic.  However, despite the weak value of $G$, one obtains a pronounced sharp feature near $\omega = \Omegam$ associated with the phonon-like $-$ polariton band.  Figs.~\ref{fig:FPresonances}(b)-(c) show zoom-ins of this feature.  One sees well-defined Fabry-Perot resonances associated with this extremely narrow $-$ band.  Remarkably, we see that these resonant features are {\it much sharper} in frequency than the intrinsic photon loss rate $\kappa_0$.
We thus have one of our key results:  phonon-like polariton modes can yield extremely sharp (width $\Delta \omega \ll \kappa_0$), large-magnitude transmission resonances, even if the optomechanical interaction is weak.  

For weak internal dissipation, the energies in the $-$ polariton band giving resonances can be easily found by combining Eqs.~(\ref{eq:polaritonenergy}) and (\ref{eq:resonancek}).  For small $G$, we can simply expand in $G$ to describe all resonances except those at the very bottom of the band.  The spacing of the resonances
is then well approximated as:
\begin{eqnarray}
	\Delta E_{-,m} & \equiv&
		|E_{-,m} -  E_{-,m+1}| \nonumber \\
	& \simeq &
			\frac{2 G^2}{\left( \delta_0 + 2 J ( 1 - \cos k_m) \right)^2 } J \sin k_m \cdot \frac{\pi}{N}
\end{eqnarray}
where $\delta_0$ is the detuning between the bare photon and phonon band at $k=0$, c.f.~Eq.~(\ref{eq:delta_k}).
Thus, for resonances in the middle of the band, we have $\Delta E_{-,m} \sim G^2 / (J N)$, where for resonances near the edge of the band, we have an even smaller scale, $\Delta E_{-,m} \sim m G^2 / (J N^2)$.  

Given the extremely small frequency separation of these resonances, one might think that even a tiny amount of intrinsic photon damping $\kappa_0$ would be enough to completely suppress them.  As seen dramatically in Figs.~\ref{fig:FPresonances}(b)-(c), this is not the case:  the resonances remain sharp even though their separation is much smaller than $\kappa_0$.
The explanation is simply that the majority of the resonances are due to polaritons which are mostly phonon-like, and hence only weakly see the photon loss rate $\kappa_0$.  

To make this more quantitative, note that the dominant effect of intrinsic dissipation is to make the effective wavevector $q[\omega]$ in Eq.~(\ref{eq:FPtransmission}) complex.  From the denominator of Eq.~(\ref{eq:FPtransmission}), this means that perfect interference between subsequent reflections will be impossible.  
By insisting that this deviation from perfect constructive interference in Eq.~(\ref{eq:FPtransmission}) is $\ll 1$, we can derive a quantitative condition for how weak the internal dissipation has to be to see the Fabry-Perot resonances in the $-$ polariton band.  Consider the relevant case $\kappa_{\textrm{ex},j} \gg J |\sin q|$, where the transmission coefficients from the probe waveguides to the lattice $\tilde{t}_j$ in Eq.~(\ref{eq:FPtransmission}) have a small magnitude, and the reflection coefficients $\tilde{r}$ have magnitude near $1$.  In this case, the condition becomes:
\begin{eqnarray}
	\left(    N \alpha  =  N \frac{\Gamma_{\rm eff}}{v_g} \right)   \ll 
		\left( |\tilde{t}|^2 =  \frac{J |\sin(k_m)| }{ \sqrt{  \kappa_{{\textrm ex},1}\kappa_{{\textrm ex},N+1} } } \right)
\end{eqnarray}
Here $\Gamma_{\rm eff}$ is the effective damping rate of the resonant polariton, and $v_g$ is its group velocity.

We can further simplify this condition by focusing on resonances away from the bottom of the band, where a small-$G$ expansion of the $-$ polariton dispersion relation 
is possible.  For such resonances, the group velocity $v_g \sim (G^2 / J) \sin k_m$, while the effective damping rate $\Gamma_{\rm eff}$ is just the sum of the mechanical
damping rate $\gamma$ and the optical damping rate $(G^2 / J^2) \kappa_0$. Taking  $\kappa_{{\textrm ex},1} = \kappa_{{\textrm ex}, N+1}=\kappa_\text{ex}$ for simplicity, our condition becomes:
\begin{eqnarray}\label{eq:resonance_condition}
	\max \left( \gamma, \frac{G^2}{J^2} \kappa_0 \right)
		\ll 
			\left( \frac{G^2}{\kappa_{\rm ex}} \right)
			\left\{
			\begin{array}{ll}
				\left( \frac{m^2}{N^3} \pi^2\right)   
					& \mbox{ if $ m \ll N$}; \\
				1/N	& \mbox{ if $ m \sim N/2$ }.
			\end{array} \right.
\end{eqnarray}	
Fig.\ \ref{fig:FPresonances} corresponds to the case $\gamma< \frac{G^2}{J^2} \kappa_0$. One can easily verify that in the opposite case $\gamma>\frac{G^2}{J^2} \kappa_0$, the resonance peaks with width much narrower than $\kappa_0$ can also be resolved only if Eq.(\ref{eq:resonance_condition}) is satisfied. We thus see that while $\kappa_0$ needs to be sufficiently small to get well-defined resonances, the relevant condition is in general {\it not} as stringent as requiring that $\kappa_0$ be smaller than the spacing of the resonances.  

We note that the interplay of OMIT and Fabry-Perot interference discussed here occurs in a similar fashion in a finite optomechanical lattice.  One could imagine two distinct situations.  The first is where the probe waveguides are coupled to sites in the interior of the lattice.  In this case, the results for the infinite lattice are recovered as long as $\kappa_{\rm ex}$ is sufficiently large; the additional reflections possible at the ends of the lattice play a minor role, as the probability of being transmitted past the sites coupled to probe waveguides is weak.  

The second situation is where the probe waveguides are coupled to the end sites of the lattice.  This case is treated in detail in Appendix \ref{appendix:Finite}.  As shown in that section, one obtains a qualitatively similar transmission spectrum to that in Fig.\ref{fig:FPresonances}, but now for a finite array (keeping other parameters the same).  On a qualitative level, the important physics here is the possibility of polariton reflections at sites $1$ and $N$.  In the infinite lattice case, these are only due to the impedance mismatch caused by the coupling to the probe waveguide; in the finite lattice, one also has reflections from the end of the lattice.  This difference does not play a large role in the qualitative nature of the interplay we discuss.


\section{Conclusion}

We have presented a study of OMIT-type effects in a one dimensional optomechanical lattice with photon-hopping
between adjacent sites, and with a strong control laser applied to each site to enhance the optomechanical interaction.
Similar to single-cavity OMIT, we find that the photon density of states can be strongly suppressed by the 
many-photon optomechanical interaction $G$, even in regimes where it is extremely weak compared to the intrinsic photon loss rate
$\kappa_0$; we discussed how these effects could be measured via the local reflection of probe photons.  

We also studied the transport of a beam of probe photons through the lattice, showing both that phonon-like polariton modes can be surprisingly effective in transport, and that one can have an interesting interplay between OMIT-style effects and geometric resonances in the lattice.  One finds the interesting result that phonon-like polaritons
can give rise to high-magnitude transmission resonances whose spectral width is much narrower than the intrinsic photon loss rate $\kappa_0$.

The general physics we describe could play an important role when one now tries to understand the effects of disorder in the OMIT lattice, as disorder could also give rise to geometric resonances.  Our work also demonstrates the importance of viewing an optomechanical lattice as a driven, dissipative system; as we have stressed, the OMIT-style interference effects that arise are a direct consequence of the structured dissipation in our system, and cannot be described by simply energy-broadening the results expected for the coherent (dissipation-free) system.

We would like to thank Marc-Antoine Lemonde for useful discussions. This work is supported by NSERC and the DARPA ORCHID program under
a grant from the AFOSR. W.C. would like to thank the Institute of Advanced Study at Tsinghua University for their hospitality during the 
writing stage of this paper.

\appendix

\section{Heisenberg-Langevin equations }
\label{appendix:EOM}

In this appendix, we have a brief derivation of the equations of motion 
for the photon and phonon fields in momentum space under 
the rotating wave approximation. The retarded GF in Eq.(\ref{eq:GFkspace}) of cavity photons in $k$ space 
 then follows from the equations of motion (EOM) of the photon and phonon fields. 

The RWA Hamiltonian in $k$ space without coupling to the bath is shown
 in Eq.~(\ref{eq:Hkspace}). The coupling to the bath is treated as
  in  the standard input-output theory. The equations of motion of
   the photon and phonon fields after this treatment of the baths are 
\begin{eqnarray}\label{eq:EOM}
i\frac{\partial \hat{d}_k}{\partial t}=\left(\omega_k-i\frac{\kappa_0}{2}\right) \hat{d}_k+G\hat{b}_k-i\sqrt{\kappa_0}\hat{\xi}_k, \nonumber\\
i\frac{\partial \hat{b}_k}{\partial t}=\left(\Omega_m-i\frac{\gamma}{2}\right) \hat{b}_k+G\hat{d}_k-i\sqrt{\gamma}\hat{\eta}_k,\nonumber\\
\end{eqnarray}
where $\hat{\xi}_k$ and $\hat{\eta}_k$ are the noise operators
of the baths which the photon and phonon modes couple to respectively.

Denoting $\hat{\phi}_k=
\left(
\begin{array}{c}
  \hat{d}_k  \\
   \hat{b}_k
\end{array}
\right)
$, the retarded GF in $k$ space $\mathbf{D}^R(k; t,t')\equiv-i\theta(t-t')\langle[\hat{\phi}_k(t), \hat{\phi}^\dag_k(t')]\rangle$ satisfies 
\begin{equation}\label{eq:GFequation}
(i\partial_t -\mathbf{M}_k)G^R(k ; t,t')=\delta(t-t'),
\end{equation}
where $\mathbf{M}_k=\left(
\begin{array}{ccc}
  \omega_k-i\frac{\kappa_0}{2}  &  G   \\
  G  &  \Omega_m-i\frac{\gamma}{2} 
\end{array}
\right)
$ obtained from the EOM in Eq.(\ref{eq:EOM}).

Fourier transforming Eq.(\ref{eq:GFequation}) to frequency domain,
 one gets the retarded GF in frequency domain as
\begin{equation}
\mathbf{D}^R(k,\omega)=(\omega \mathbf{1}-\mathbf{M}_k)^{-1}.
\end{equation}
The retarded GF $\mathbf{D}^R(k,\omega)$ obtained above is a $2\times 2$ matrix. 
The $(1,1)$ component gives the retarded GF for the photon component $D^R_c(\omega, k)$ in Eq.(\ref{eq:GFkspace}).

\section{Calculation of photon Green function in the presence of external waveguides}
\label{appendix:GFCalc}

In this appendix, we present the calculation to obtain the full retarded Green's function of cavity photons 
in Eq.(\ref{eq:GFlocaldressed}) and Eq.(\ref{eq:GFdressed}) taking into account the effect of coupling to the probe waveguide.

We first get the local retarded Green function of cavity photons in Eq.(\ref{eq:GFlocaldressed}) dressed by probe coupling. 

The coupling to probe waveguide on site $1$ for reflection experiments introduces an effective imaginary potential  to site $1$, 
which in the position representation is 
\begin{equation}
V(1,1)=-i\frac{1}{2}\kappa_\text{ex,1}
\end{equation}
and vanishes on all other sites. The bare retarded Green's function of cavity photons in real space
 $D^R_\text{0,c}(\omega;1,1)$ is obtained from Eq.(\ref{eq:photonGF}) to be
\begin{equation}
D^R_\text{0,c}(\omega;1,1)=\frac{1}{2iJ\sin{q[\omega]}}.
\end{equation}

From Dyson's equation, the full local retarded Green's function of cavity photons can be obtained as
\begin{eqnarray}
&&D^R_\text{c}(\omega;1,1)\nonumber\\
&&\ \ =D^R_\text{0,c}(\omega;1,1)+D^R_\text{0,c}(\omega;1,1)V(1,1)D^R_\text{c}(\omega;1,1).\nonumber\\
\end{eqnarray}
The full retarded Green Function $D^R_\text{c}(\omega;1,1)$ is then
\begin{equation}
D^R_\text{c}(\omega;1,1)=\frac{1}{(D^R_\text{0, c}(\omega;1,1))^{-1}-V(1,1)},
\end{equation}
as obtained in Eq.(\ref{eq:GFlocaldressed}).

For the transmission experiments, two probe waveguides are introduced on site $1$ and site $N+1$ respectively.
 The couplings to the waveguides introduce an effective imaginary potential to site $1$, i.e., $V(1,1)=-i\frac{1}{2}\kappa_\text{ex,1}$, 
 and site $N+1$, i.e., $V(N+1,N+1)=-i\frac{1}{2}\kappa_\text{ex,N+1}$ respectively.
  These potentials can be written as a matrix in the space spanned by states $|1\rangle$ 
  and $|N+1\rangle$ as
  
\begin{eqnarray}
&&{\bf V}=\left(
\begin{array}{ccc}
  V(1,1) &  V(1,N+1)   \\
  V(N+1,1)  &  V(N+1,N+1)   \\
\end{array}
\right)\nonumber\\
&&\ \ \ \ \ \ \ \ \ \ \ \ \ =\left(
\begin{array}{ccc}
 -i\frac{1}{2}\kappa_\text{ex,1} &  0 \\
 0  & -i\frac{1}{2}\kappa_\text{ex,N+1}   \\
\end{array}
\right).
\end{eqnarray}

The Full retarded Green's functions of cavity photons satisfy the Dyson's equation
\begin{widetext}
\begin{eqnarray}\label{eq:fullGF}
&&\left(
\begin{array}{ccc}
  D^R_\text{c}(\omega;1,1) & D^R_\text{c}(\omega;1,N+1)  \\
 D^R_\text{c}(\omega;N+1,1)  &  D^R_\text{c}(\omega;N+1,N+1)   \\
\end{array}
\right)=\left(
\begin{array}{ccc}
  D^R_\text{0,c}(\omega;1,1) & D^R_\text{0,c}(\omega;1,N+1)  \\
 D^R_\text{0,c}(\omega;N+1,1)  &  D^R_\text{0,c}(\omega;N+1,N+1)   \\
\end{array}
\right)\nonumber\\
&&\ \ \ \ \ \ \ \ \ \ \ \ \ \ \ +\left(
\begin{array}{ccc}
  D^R_\text{0,c}(\omega;1,1) & D^R_\text{0,c}(\omega;1,N+1)  \\
 D^R_\text{0,c}(\omega;N+1,1)  &  D^R_\text{0,c}(\omega;N+1,N+1)   \\
\end{array}
\right){\bf V}\left(
\begin{array}{ccc}
  D^R_\text{c}(\omega;1,1) & D^R_\text{c}(\omega;1,N+1)  \\
 D^R_\text{c}(\omega;N+1,1)  &  D^R_\text{c}(\omega;N+1,N+1)   \\
\end{array}
\right).\nonumber\\
\end{eqnarray}
\end{widetext}

The bare retarded Green's function of cavity photons $D^R_\text{0,c}(\omega;N,1)$ is given by Eq.(\ref{eq:photonGF}). 
Combined with Eq.(\ref{eq:fullGF}), the full retarded Green's function of cavity photons $D^R_{c}(\omega;N+1,1)$ 
can be obtained as shown in Eq.(\ref{eq:GFdressed}).

\


\section{Photon Green function in a finite optomechanical array}
\label{appendix:Finite}

 \begin{figure*}[htbp]
	\includegraphics[width= \textwidth]{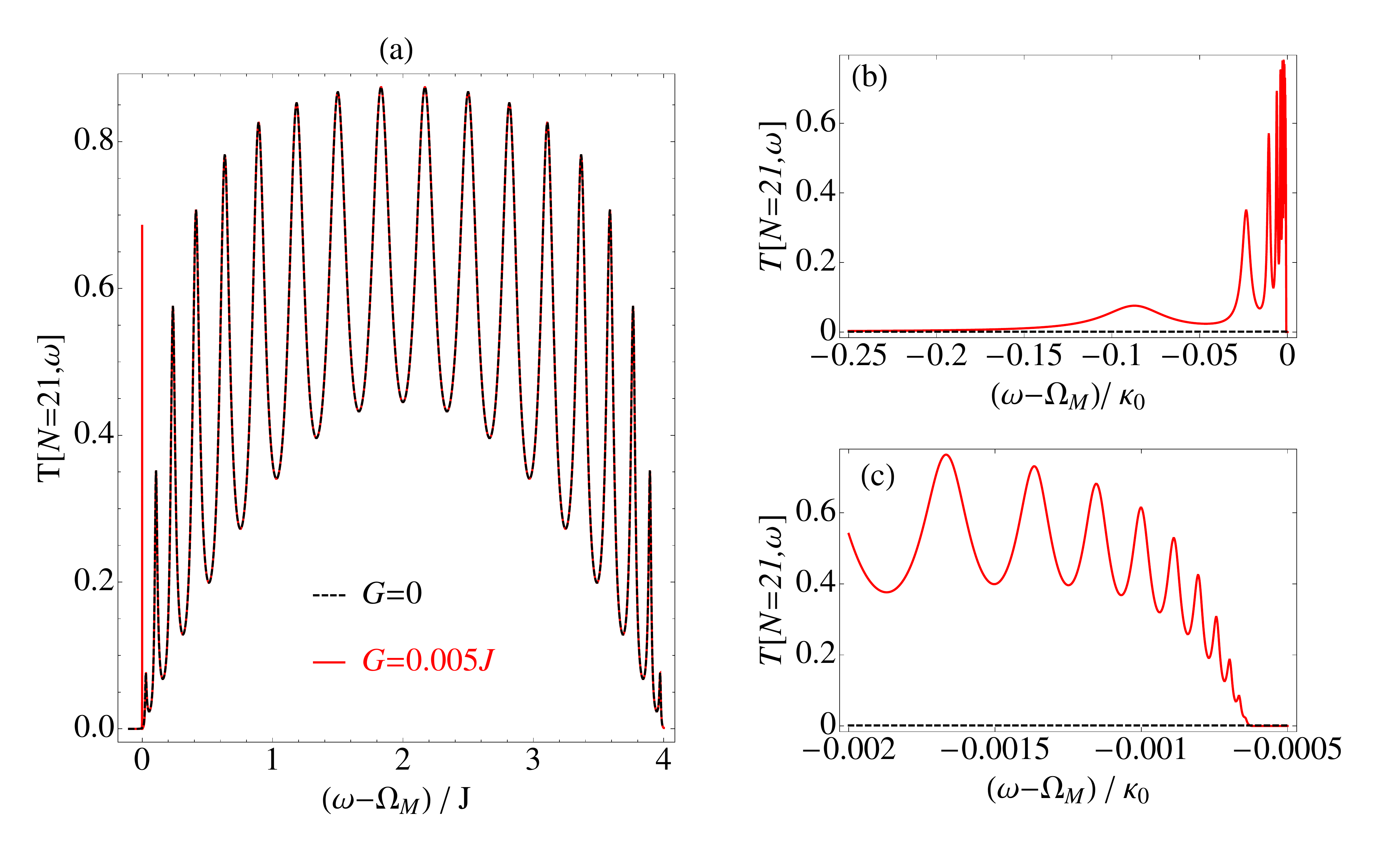}
	\caption{
		(color online) Transmission probability for probe light from sites $j=1$ 
		to $j=21$ for a finite optomechanical array of $21$ cells, as a function of the probe beam frequency $\omega$.
		  (a) Black curve: transmission spectrum without
		optomechanical coupling, $G=0$. Red curve: transmission spectrum with $G = \kappa_0/2$. 
		 All the other parameters are the same as in Fig.\ref{fig:FPresonances}.
		  (b), (c)  Zoom in of the $-$ polariton band features in (a).  
		One again sees Fabry-Perot style resonances. The location and width of the Fabry-Perot resonances are 
		almost the same as in the infinite array in Fig.\ref{fig:FPresonances}, whereas the resonance heights are different. } 
	\label{fig:FPresonance_finite}
\end{figure*}

In this appendix, we present the retarded Green's function of cavity photons in a finite optomechanical array with $N+1$ cells. We will 
show that the Fabry-Perot interference also occurs in a finite optomechanical array
 of $N+1$ cells with probe coupling on the two end sties.  In the limit $\kappa_\text{ex,1}, \kappa_\text{ex, N+1}\gg J$
  the full retarded Green's functions $D^R_c(\omega;N+1,1)$ for the infinite array and finite array 
  reduce to  the same form so the transmission properties
  are the same in the two systems in this limit. 

The retarded Green's function of cavity photons in a finite optomechanical array can be obtained from
 the bare retarded Green's function in an infinite array following the same procedure in Appendix A. 

We first obtain the retarded Green's function in a semi infinite array by introducing an effective potential 
to cut the hopping between site $0$ and site $1$ in an infinite array. In the position representation 
spanned by the position states $|0\rangle$ and $|1\rangle$,
the effective potential can be written as
\begin{eqnarray}
&&{\bf V}=\left(
\begin{array}{ccc}
  V(0,0) &  V(0,1)   \\
  V(1,0)  &  V(1,1)   \\
\end{array}
\right) =\left(
\begin{array}{ccc}
0&  J \\
 J  & -i\frac{1}{2}\kappa_\text{ex,1}   \\
\end{array}
\right).\nonumber\\
\end{eqnarray}
Here $V(1,1)=-i\frac{1}{2}\kappa_\text{ex,1} $ takes into account the external decay rate due to coupling with the probe waveguide 
on site $1$. 

The retarded Green's function $D^R_\text{s,c}(\omega;n,n')$ in a semi-infinite optomechanical array 
can be obtained from the Dyson's equations below
\begin{widetext}
\begin{eqnarray}\label{eq:Dysonequation}
&& \left(
\begin{array}{ccc}
  T(0,0) &  T(0,1)   \\
  T(1,0)  &  T(1,1)\\
\end{array}
\right)
= \left(
\begin{array}{ccc}
  V(0,0) &  V(0,1)   \\
  V(1,0)  &  V(1,1)\\
\end{array}
\right)\nonumber\\
&&\ \ \ \ \ \ \ \ \ \ +\left(
\begin{array}{ccc}
  V(0,0) &  V(0,1)   \\
  V(1,0)  &  V(1,1)\\
\end{array}
\right)\left(
\begin{array}{ccc}
  D^R_{0,c}(\omega;0,0) & D^R_{0,c}(\omega;0,1)  \\
 D^R_{0,c}(\omega;1,0)  &  D^R_{0,c}(\omega;1,1)   \\
\end{array}
\right) \left(
\begin{array}{ccc}
  T(0,0) &  T(0,1)   \\
  T(1,0)  &  T(1,1)\\
\end{array}
\right);\nonumber\\
&&\left(
\begin{array}{ccc}
  D^R_{s,c}(\omega;n,n) & D^R_{s,c}(\omega;n,n')  \\
 D^R_{s,c}(\omega;n',n)  &  D^R_{s,c}(\omega;n',n')   \\
\end{array}
\right)=\left(
\begin{array}{ccc}
  D^R_{0,c}(\omega;n,n) & D^R_{0,c}(\omega;n,n')  \\
 D^R_{0,c}(\omega;n',n)  &  D^R_{0,c}(\omega;n',n')   \\
\end{array}
\right)\nonumber\\
&&\ \ \ \ \ \ \ \ \ \ +\left(
\begin{array}{ccc}
 D^R_{0,c}(\omega;n,0) & D^R_{0,c}(\omega;n,1)  \\
 D^R_{0,c}(\omega;n',0)  &  D^R_{0,c}(\omega;n',1)   \\
\end{array}
\right) \left(
\begin{array}{ccc}
  T(0,0) &  T(0,1)   \\
  T(1,0)  &  T(1,1)\\
\end{array}
\right)\left(
\begin{array}{ccc}
  D^R_{0,c}(\omega;0,n) & D^R_{0,c}(\omega;0,n')  \\
 D^R_{0,c}(\omega;1,n)  &  D^R_{0,c}(\omega;1,n')   \\
\end{array}
\right).\nonumber\\
\end{eqnarray}
\end{widetext}

We get the retarded Green's function for a semi-infinite optomechanical array as
\begin{equation}
D^R_{s,c}(\omega;n,n')=\frac{1}{2iJ\sin{q[\omega]}}\left(e^{iq[\omega]|n-n'|}-e^{iq[\omega]|n+n'|}\zeta_1\right),
\end{equation}
where $q[\omega]$ is the same as defined in Eq.(\ref{eq:k}) and $\zeta_1$ is defined as
\begin{equation}
\zeta_1=\frac{1-i\frac{\kappa_\text{ex,1}}{2}e^{-iq[\omega]}}{1-i\frac{\kappa_\text{ex,1}}{2}e^{iq[\omega]}}.
\end{equation}

To get the retarded Green's function for the finite array with $N+1$ cells, we follow the same idea as above and introduce an 
effective potential to cut the hopping between site $N+1$ and site $N+2$ as
\begin{eqnarray}
{\bf V}&=&\left(
\begin{array}{ccc}
  V(N+1,N+1) &  V(N+1,N+2)   \\
  V(N+2,N+1)  &  V(N+2,N+2)   \\
\end{array}
\right) \nonumber\\
&=&\left(
\begin{array}{ccc}
 -i\frac{1}{2}\kappa_\text{ex, N+1}   &  J \\
 J  & 0 \\
\end{array}
\right).\nonumber\\
\end{eqnarray}
Here $V(N+1,N+1)=-i\frac{1}{2}\kappa_\text{ex, N+1} $ takes into account the external decay rate due to coupling with the probe 
waveguide on site $N+1$. 

Following the same procedure in Eq.(\ref{eq:Dysonequation}), we get the retarded Green's function of cavity photons
 in a finite array of $N+1$ cells as 
\begin{eqnarray}\label{eq:GFfinite}
&&D^R_{c,f}(\omega;N+1,1)=\nonumber\\
&&\ \ \ \frac{1}{2i\sin{q[\omega]}}\frac{(-1+\zeta_1Z^2)(-\zeta_{N+1}+1/Z^2)}{\zeta_1\zeta_{N+1}Z^{N+2}-1/Z^{N+2}},\nonumber\\
\end{eqnarray}
where $Z\equiv e^{iq[\omega]}$ and $\zeta_{N+1}$ is defined as 
\begin{equation}
\zeta_{N+1}=\frac{1-i\frac{\kappa_\text{ex, N+1}}{2}e^{-iq[\omega]}}{1-i\frac{\kappa_\text{ex, N+1}}{2}e^{iq[\omega]}}.
\end{equation}

One can easily verify that Eq.(\ref{eq:GFfinite}) and Eq.(\ref{eq:GFdressed}) share the same denominator of Eq.(\ref{eq:GFdressed}).
 Since the Fabry-Perot interference is manifested by the same parameter $Z^{2N}-1$ in the denominator,
  the two systems show very similar properties of Fabry-Perot interference, 
  i.e., the resonance frequencies and width of the transmission spectrum in the two systems are the same,
  though the peak heights of the two systems are different due to the different numerators of the Green's functions 
  in Eq. (\ref{eq:GFfinite}) and Eq.(\ref{eq:GFdressed}). For comparison, the transmission spectrum
   from the first site to the last site of a finite array of $N+1=21$ cells is 
  shown in Fig.\ref{fig:FPresonance_finite}, with the same parameters as in Fig.\ref{fig:FPresonances}.

In the limit $\kappa_\text{ex,1}, \kappa_\text{ex, N+1}\gg J$, one can Taylor expand Eq.(\ref{eq:GFfinite}) and 
Eq.(\ref{eq:GFdressed}) treating $J/\kappa_\text{ex,1}$ and $J/\kappa_\text{ex, N+1}$ as small parameters 
and the retarded Green's functions of the two systems reduce to the same form
\begin{equation}
D^R_c(\omega;N+1,1)=\frac{8JZ^{N}\sin{q[\omega]}}{(Z^{2N}-1)\sqrt{\kappa_\text{ex,1}\kappa_\text{N+1,ex}}}.
\end{equation}
In this limit, the two systems are equivalent and the transmission properties from site $1$ to site $N+1$ are the same.

\bibliography{OMIT}

\end{document}